\documentclass[reprint,superscriptaddress,twocolumn]{revtex4}

\usepackage{xcolor}
\usepackage{color}
\usepackage{soul}
\usepackage{graphicx}
\usepackage{dcolumn}
\usepackage{bm}
\usepackage{ulem}
\usepackage{amsmath}
\usepackage{amsfonts}

\begin{document}

\preprint{APS/123-QED}

\title{A robust basis for multi-bit optical communication with vectorial light}

\author{Keshaan Singh}
\author{Isaac Nape}%
\author{Wagner Tavares Buono}
\author{Angela Dudley}
\author{Andrew Forbes}
 \email{andrew.forbes@wits.ac.za}
\affiliation{School of Physics, University of the Witwatersrand, Private Bag 3, Johannesburg 2050, South Africa}%

\date{\today}

\begin{abstract}\noindent Increasing the information capacity of communication channels is a pressing need, driven by growing data demands and the consequent impending data crunch with existing modulation schemes. In this regard, mode division multiplexing (MDM), where the spatial modes of light form the encoding basis, has enormous potential and appeal, but is impeded by modal noise due to imperfect channels.  Here we overcome this challenge by breaking the existing MDM paradigm of using the modes themselves as a discrete basis, instead exploiting the polarization inhomogeneity (vectorness) of vectorial light as our information carrier.  We show that this encoding basis can be partitioned and detected almost at will, and measured in a channel independent fashion, a fact we confirm experimentally using atmospheric turbulence as a highly perturbing channel example.  Our approach replaces conventional amplitude modulation with a novel modal alternative for potentially orders of magnitude channel information enhancement, yet is robust to fading even through noisy channels, offering a new paradigm to exploiting the spatial mode basis for optical communication.

\end{abstract}

\maketitle



\begin{figure}[t]
    \centering
    \includegraphics[width=\linewidth]{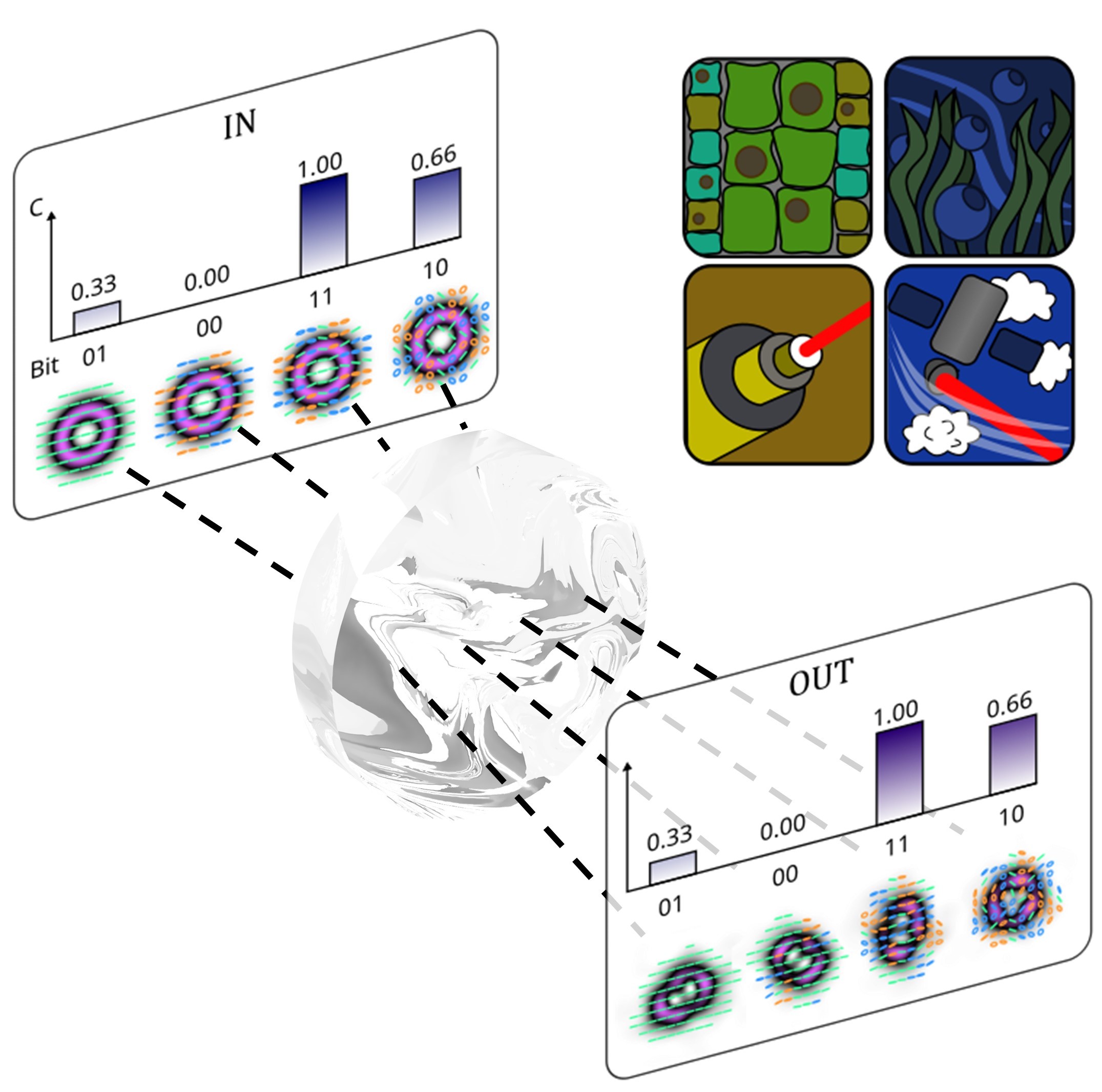}
    \caption{\textbf{Modal vectorness for multi-bit encoding.} Conceptual visualization of a sequence of vector beams with tuneable vectorness (denoted as $\mathcal{C}$). Each $\mathcal{C}$ value is binned into the user-defined basis allowing for the received $\mathcal{C}$ to map to a bit-string of length determined by the basis choice. After propagating through an aberrating channel (e.g. atmospheric turbulence, underwater, single-mode fiber or even transparent cellular material) the intensity and polarization profiles of the beams are distorted, however the $\mathcal{C}$ is left unaffected.}
    \label{fig:concept}
\end{figure}

\noindent Optical communication has been an integral part of human society since recorded time, initially visual and thus free-space based, with wire-based solutions only emerging 200 years ago, from copper wire to modern day fibre optics. Today, our optical communication solutions are rapidly reaching their capacity limit \cite{Richardson1,richardson2013space}, requiring new degrees of freedom for packing information into light.  Here so-called structured light comes to the fore \cite{forbes2021structured}, where light's spatial degree of freedom is used in space division multiplexing (SDM) \cite{li2014space}  and  mode division multiplexing (MDM) \cite{Berdague1982A}, for more channels and more capacity per channel, with significant advances over recent years \cite{trichili2019communicating,trichili2020roadmap}.  Topical among the multiplexing techniques is the use of orbital angular momentum (OAM) modes, with excellent reviews available on the topic \cite{padgett2017orbital,Willner2015,willner2021orbital}. For instance, in free-space reaching petabits-per-second data rates in the laboratory \cite{wang2014n} and 80 gigabits per second over 260 metres \cite{zhao2016experimental} while advances in optical fibre have shown 1.6 Tbit/s OAM communication over kilometer lengths with a custom ring core fiber \cite{Bozinivic2013}.  In both these channels, modal cross-talk is a limiting factor.

Atmospheric turbulence induced modal scattering in free-space links is a particularly severe example \cite{JaimeA.Anguita2008,ren2016experimental,krenn2014communication,rodenburg2012a,zhang2020mode,malik2012influence,chen2016changes,tyler2009influence,cox2020structured}, impeding classical and quantum communication links with structured light to only hundreds of meters \cite{zhao2016experimental}. To mitigate this, there has been active research in searching for robust states of structured light in turbulence, including Bessel-Gaussian~\cite{mphuthi2018bessel,mphuthi2019free,lukin2014mean,bao2009propagation,zhu2008propagation,nelson2014propagation,ahmed2016mode,cheng2016channel,doster2016laguerre,watkins2020experimental,vetter2019realization,yuan2017beam}, Hermite-Gaussian~\cite{cox2019hglg,ndagano2017c,Restuccia2016,ndagano2017comparing}, Laguerre-Gaussian~\cite{Trichili2016,zhao2015capacity,zhou2019using,xie2016experimental,li2017power} and Ince-Gaussian~\cite{krenn2019turbulence} beams, but none have shown to be robust in laboratory and real-world experiments.  This can be explained best from the perspective of OAM: the atmosphere itself can transfer OAM to and from the beam, so that all beams will be affected equivalently regardless of their initial structure \cite{klug2021orbital}.  A recent development has been the creation of vectorial combinations of OAM, for vector vortex light \cite{zhan2009cylindrical,rosales2018review} with inhomogeneous polarization patterns, and these too have found applications in optical communication, both classical \cite{Milione2015d,zhu2021compensation} and quantum \cite{ndagano2017deterministic,Sit2017,nape2018self,ndagano2018creation}.  Unfortunately they have been found not to be resilient in turbulence \cite{cai2008average,ji2010propagation,wang2008propagation,Cox:16,Ndagano2017}, a fact that has been generalised to many different channels, revealing that the vectorial nature of such light remains invariant even as the spatial pattern itself alters \cite{nape2022revealing}.  It is this ``altering'' or distorting of the pattern that impedes MDM, since its very premise is to ``recognise'' the original pattern in its undistorted form.

Here we present a novel approach to exploiting MDM in optical communication, foregoing the discrete modal basis of MDM and instead exploiting the polarisation inhomogeneity (vectorness) of vectorial light as an encoding basis in a manner that does not require the mode itself to be recognised: a modal basis without the penalty of a modal detection. Similar to amplitude, it spans 0 (scalar light) to 1 (fully vectorial) but with the pertinent benefit that the full range can be used as a high-dimensional alphabet, rather than the two dimensional (on/off) amplitude alphabet.  Because the vectorness is an invariant quantity, it remains intact even in the presence of aberrations, where all parties will receive the same information regardless of their particular channel conditions. This implies the potential for optical communication free of modal-induced noise without the need for adaptive optics or digital corrective procedures. We demonstrate our encoding scheme on a highly-aberrated channel of dynamically changing atmospheric turbulence, chosen as an extreme example of a noisy channel, showing correction-free data transmission under a wide range of conditions, with the dimensionality of the encoding primarily limited by detector noise. Our approach replaces conventional amplitude modulation with a modal alternative for potentially orders of magnitude channel information enhancement, offering a new approach to exploiting structured light for optical communication.   

\section*{RESULTS}

\begin{figure*}[t]
    \centering
    \includegraphics[width=0.8\linewidth]{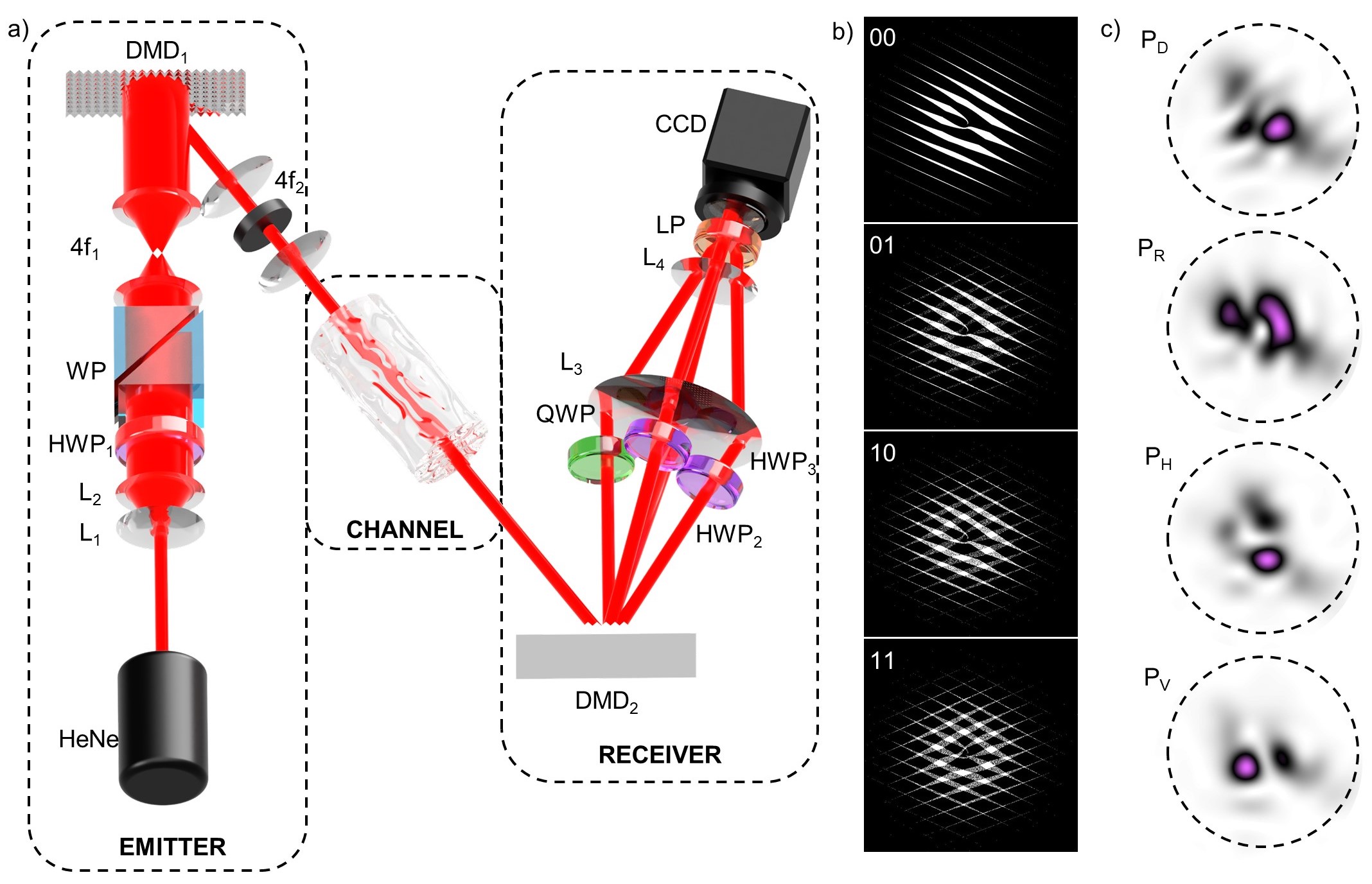}
    \caption{\textbf{Experimental demonstration.} a) Diagram showing the setup used to generate vector fields with tune-able concurrence (EMITTER) and measure the concurrence (RECEIVER) after propagating through heated air (CHANNEL) - [L$_i$ - lens, HWP$_i$ - half-wave plate, WP - Wollaston prism, DMD$_i$ - digital micro-mirror device, LP - linear polarizer, CCD - charge-coupled device]. b) Exemplar multiplexed binary amplitude holograms used to holographically control the concurrence of generated beams via weighting parameter $a$ - associated bit strings are displayed. c) Exemplar Stokes intensities which are integrated over the dashed circles in order to calculate the concurrence [$P_H$ - horizontal, $P_V$ - horizontal, $P_D$ - diagonal and $P_R$ - right-circular polarization integrated powers].}
    \label{fig:setup}
\end{figure*}

\vspace{0.5cm}
\noindent \textbf{A new encoding scheme.} Our technique involves exploiting the spatial mode basis using vector fields of the form

\begin{equation}
    |\Psi\rangle = \sqrt{a}|\psi_+\rangle|H\rangle + \sqrt{1-a}|\psi_-\rangle|V\rangle\,,
    \label{eqn:VB}
\end{equation}

where $|\psi_{\pm}\rangle$ represents any pair of orthogonal spatial modes, $|H(V)\rangle$ are the horizontal(vertical) linear polarization Jones vectors and the real amplitude determined by $a$ ensures that the total power in the field remains constant for $a\in[0,1]$.  Such beams are ubiquitous, and in particular are the natural modes of both optical fibre and free space when $|\psi_{\pm}\rangle$ are chosen appropriately \cite{rosales2018review}.  It has been shown that fields of this type have a degree of non-separability in their spatial and polarization degrees-of-freedom (DoFs) which can be quantified using a quantum inspired metric (the concurrence), termed the vector quality factor \cite{mclaren2015,ndagano2016beam}, and which we will refer to as the \textit{vectorness} for short. The concurrence, $\mathcal{C}$, for the field in Eqn \ref{eqn:VB} is given by \cite{wootters1998entanglement} 

\begin{equation}
    \mathcal{C} = 2\sqrt{\langle\psi_+|\psi_+\rangle\langle\psi_-|\psi_-\rangle - |\langle\psi_+|\psi_-\rangle|^2}\,,
    \label{eq:con}
\end{equation}

\noindent which reduces to $\mathcal{C}=2|\sqrt{a(1-a)}|$. We see that the weighting parameter $a$ allows $\mathcal{C}$ to be varied monotonically from a minimum value of $\mathcal{C}=0$, which represents completely scalar fields (homogeneously polarized), to a maximum of $\mathcal{C}=1$ which represents fields with maximally non-separable spatial and polarization DoFs (inhomogeneously polarized), with the required amplitude modulation trivially found from

\begin{equation}
    a = \frac{1}{2}(1 \pm \sqrt{1 - \mathcal{C}^2})\,.
    \label{eqn:a_eqn}
\end{equation}

\noindent Note that both $C$ and $a$ span from 0 to 1, but the invariance of $C$ allows us to use it as a multi-bit encoding basis. This is because all measures of $C$ will agree on the value regardless of the detector type (and we will show, regardless of the channel conditions), a property inherited from its quantum origin.  This is in stark contrast to a measurement of the amplitude, $a$, directly, where disparate detector efficiencies and channel noise/loss mean no universal agreement on $a$, thus reducing the full multi-bit bandwidth spanning 0 through to 1, to just 0 or 1 (one bit) for light or no light. To use the vectorness as measured by $\mathcal{C}$ as an encoding basis we can assign unique information to values separated by $\Delta C$, for $N \approx 1/\Delta C$ as the number of elements in the basis (we will return later to how small $\Delta C$ can be). This allows us to transmit $d = \log_2 N$ bits per on/off pulse, rather than the one bit with the traditional non-modal amplitude approach. We illustrate this concept in Fig.~\ref{fig:concept}, where initial modes (IN) are passed through an aberrating channel and emerge distorted (OUT). Although the spatial mode structure appears scrambled and would have high modal cross-talk, the vectorness remains intact with no cross-talk, and can therefore be used as a multi-bit encoding basis.  Here concurrence values of $\mathcal{C}=\{0, 0.33, 0.66, 1\}$ are used with $N=4$ as an example.  Although in the remainder of this paper we will use turbulence as our example channel, it could be optical fibre, underwater or cellular media too, as illustrated in Fig.~\ref{fig:concept}.

\vspace{0.5cm}
\noindent \textbf{Invariance of the basis to turbulence.} Now that we have defined how we construct a communication basis from the concurrence of classical vector beams, we can investigate how this basis responds to propagation through the atmosphere where spatially varying air densities due to both pressure and temperature variations induce a spatially varying refractive index according to the Gladstone-Dale law. We select turbulence as an example only, chosen because it represents a particularly dynamic and extreme distorting medium.  We can express the total phase change, $\Phi(\vec{r}_T)$, through a width of turbulent medium using the thin screen approximation \cite{herman1990method}. This phase acts only on the spatial DoF via the operator in the position basis $\hat{P}_\Phi=\int e^{i\Phi(\vec{r}_T)}|\vec{r}_T\rangle\langle \vec{r}_T|d\vec{r}_T$. We also choose to allow propagation to the far-field for phase-only distortions to become phase and intensity fluctuations, modelled by using a Fourier transform operator $\mathcal{F}$. We thus express our field after propagating into the far-field through turbulence as 

\begin{equation}
    |\phi_\pm\rangle = \mathcal{F}\hat{P}_\Phi|\psi_{\pm}\rangle\,.
\end{equation}

We can now take note of an important property of this transformation, namely, that it is unitary and therefore preserves inner products such that $\langle\phi_{\pm}|\phi_{\pm}\rangle = \langle\psi_{\pm}|\psi_{\pm}\rangle\ $ and $\langle\phi_{\mp}|\phi_{\pm}\rangle = \langle\psi_{\mp}|\psi_{\pm}\rangle\ $.  From Eqn~\ref{eq:con} we see immediately that this implies the invariance of the vectorness (concurrence) to this channel, and obviously extended to other similar transformations. The result of this is illustrated in Fig.~\ref{fig:concept}, where the vectorness and the bit-strings they are mapped to are unchanged through the channel, even though the intensity and polarization profiles are severely distorted. A more detailed derivation of this theory is given in the Supplementary Information. This theoretical concept reveals the potential of vectorness as a robust multi-bit information carrier.

\vspace{0.5cm}
\noindent \textbf{Practical demonstration.} In order to verify the effectiveness of the suggested encoding scheme, we utilized the experimental setup seen in Fig.~\ref{fig:setup}, with a more detailed breakdown given in the Methods section. The vector beams were generated using an interferometer consisting of a Wollaston prism (WP), an imaging system and a digital-micromirror device (DMD$_1$), where the independent complex modulation of horizontally and vertically polarized components of a diagonally polarized (approximate) plane wave was facilitated by two multiplexed binary amplitude holograms \cite{rosales2020polarisation}. In order to control $\mathcal{C}$ the relative efficiencies of the holograms were tuned according to Eqn.~\ref{eqn:a_eqn} by varying the amplitude, $a$, with exemplar multiplexed holograms for a $N=4$ concurrence basis shown in Fig.~\ref{fig:setup}.b. The vectorial beam was passed through a dynamically aberrated channel created by a heater set to a steady state temperature of $T \approx 185^\circ$C to induce turbulence in the air along $\approx200$ mm of beam path.  The resulting beam was then passed to the detection system, a custom built single shot Stokes polarimetry arrangement, composed of a second DMD$_2$ encoded with multiplexed binary gratings which diffracted $4$ copies of the vector field into the $\pm1$ diffraction orders. The beams were allowed to propagate to the far field for the phase distortion to manifest as both amplitude and phase distortion. The four copies were then imaged onto the sensor of a CCD camera - initially a low-cost FLIR Chameleon was used and later replaced by a higher quality FLIR Grasshopper to highlight the role of detector sensitivity. Prior to the CCD, each of the four copies were passed through different combinations of quarter-wave and half-wave plates and a common linear polarizer in order to capture horizontally ($I_H$), vertically ($I_V$), diagonally ($I_D$) and right-circularly ($I_R$) polarized Stokes intensity measurements from which the concurrence was calculated. Furthermore, the Stokes measurements form an over-complete set of polarisation projections, making our decoding step independent of the polarisation basis that the modes were prepared in, i.e., the sender or the channel can alter the polarization basis without any effect on the outcome.  Figure \ref{fig:setup}.c shows exemplar Stokes intensities integrated over the dashed region to obtain associated powers $P_{H,V,D,R}$, from which the concurrence could be determined (see Methods).  

\begin{figure*}[t]
    \centering
    \includegraphics[width=0.9\linewidth]{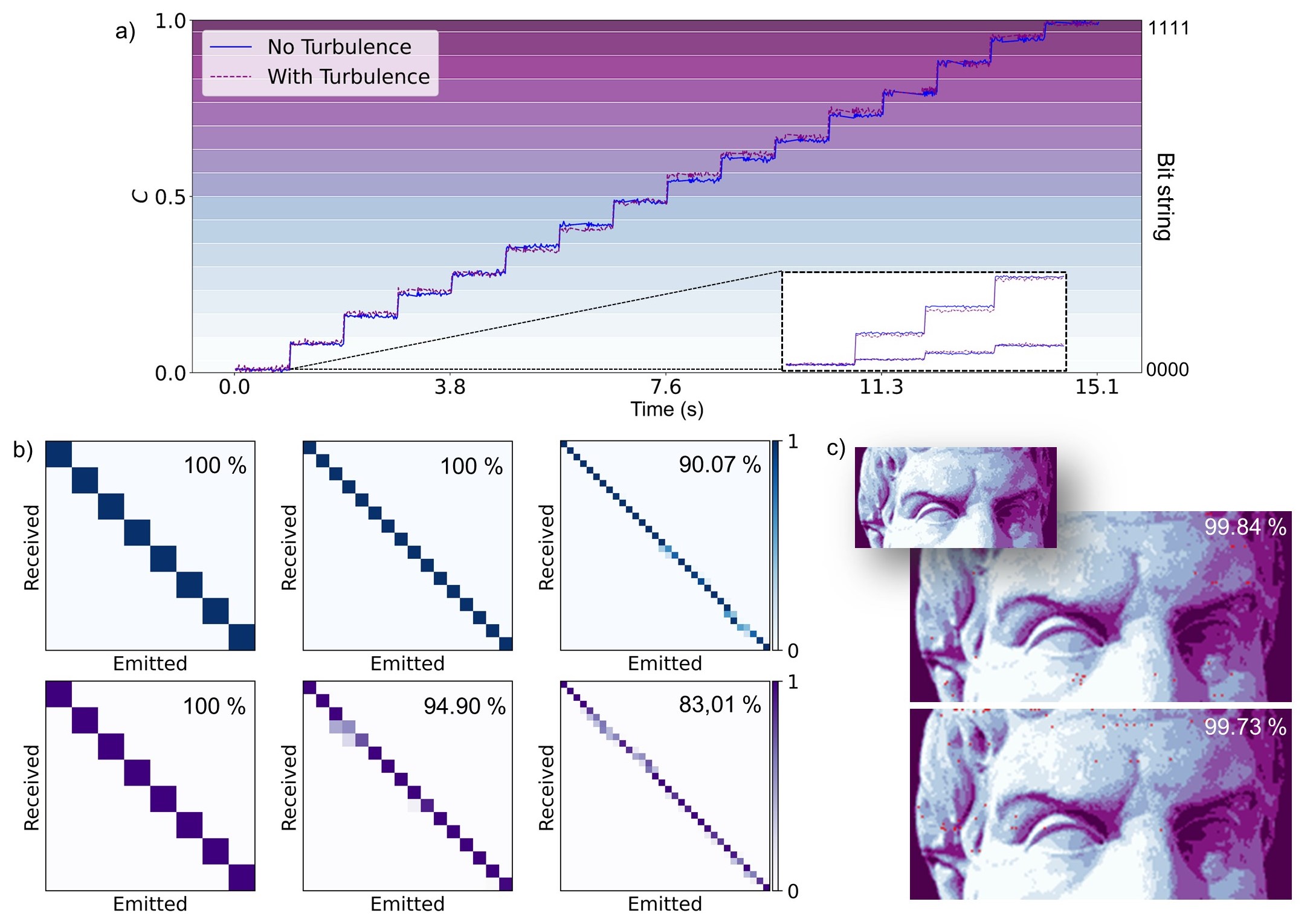}
    \caption{\textbf{Discreetized bases, crosstalk and multi-bit information transfer.} a) Received concurrence of an input value varied in 16 discreet steps (insets show partial results for 8 and 32 steps) through still (solid line) and turbulent (dashed line) air. b) Crosstalk matrices (with associated fidelity) for the emitted and received concurrence through still (top) and turbulent (bottom) air for bases with $N=8,16$ and $32$ (left to right). c) 3 Bit (8 level) image transmitted through still (top) and turbulent (bottom) air - red pixels highlight errors (inset shows the ground truth image) while the values report the image fidelity.}
    \label{fig:results}
\end{figure*}

\begin{figure*}[t]
    \centering
    \includegraphics[width=\linewidth]{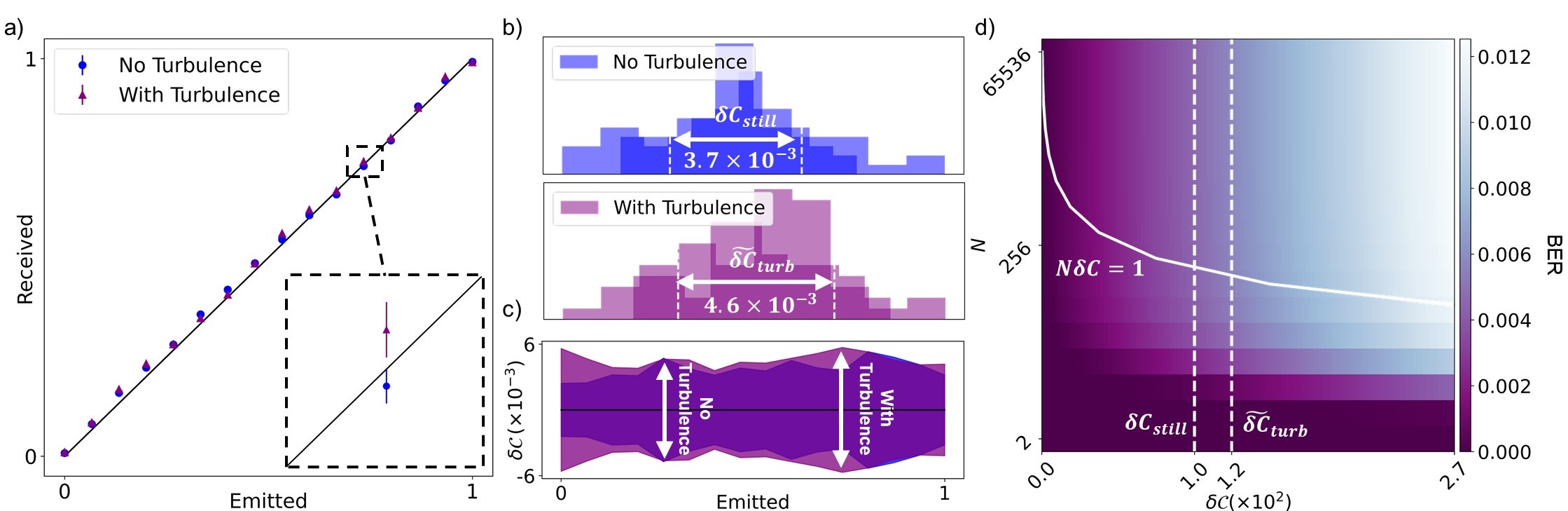}
    \caption{\textbf{Noise, statistical variation and bit-errors.} a) Mean received concurrence (over 50 points) per element in the $N=16$ basis in both still (circles) and turbulent (triangles) air, error bars are visible in the inset. b) Distribution of received $\mathcal{C}$ values about their target for two example basis elements both without (top) and with (bottom) turbulence - values report the mean deviations across the $N=16$ basis. c) Measured standard deviation for each element in the $N=16$ basis through still and turbulent air, where the large overlap illustrates the turbulence invariance. d) Theoretically predicted BER as a function of the statistical spread $\delta\mathcal{C}$ and basis choice $N$, the white solid line represents the points where $N\delta\mathcal{C}=1$ and white dashed lines represent the deviations of our receiver under different channel conditions - $\delta\mathcal{C}_\text{still}$ and $\tilde{\delta\mathcal{C}}_\text{turb}$.}
    \label{fig:limitations}
\end{figure*}
 
\vspace{0.5cm}
\noindent \textbf{Multi-bit encoding.} The key to our proposed technique is that $\mathcal{C}$ does not change due to beam distortions, so that the full range from 0 to 1 can be used in a user-defined number of steps ($N$) independent of how perturbing the channel is.  This sub-division of the available encoding space is shown in Fig.~\ref{fig:results}.a for $N=16$ (with $N=\{8,32\}$ included as an inset). To indicate the low cross-talk, the transmitted vector beam was allowed to ``idle'' at each basis element (delineated into bins shown as shaded bars) while repeated measurements were taken, confirming only small dynamic changes in still and turbulent air alike. The robust nature of the encoding is quantified by the cross-talk matrices in Fig.~\ref{fig:results}.b for up to $N=32$, corresponding to 32 vectorial modes, yet with low levels of cross-talk. Finally, we use the system to transmit information over this dynamically changing turbulent channel with no adjustment to the received data signal, resulting in the high fidelity images shown in Fig.~\ref{fig:results}.c for still (top) and turbulent (bottom) air using $N=8$.  The consistency of the results under different channel conditions validates the concept.  The notable feature of this approach is that the number of modes used can be tailored up to a maximum $N_\text{max} = 1/ \delta C$, where $\delta C$ is the inherent noise in the system - calculated as the standard deviation of repeated measurements at a given $\mathcal{C}$.  Next, we take a closer look at this to estimate the potential of the approach and show that, as expected, $\delta C$ is limited primarily by detector noise and not the condition of the channel itself.

\vspace{0.5cm}
\noindent\textbf{Noise and information capacity.} We see from the ``idling'' results of Fig.~\ref{fig:results}.a that the noise of the still and turbulent air are comparable, suggesting that detector noise is the primary cause of the statistical variation of a given vectorness about the target, as anticipated by theory. This results in only very small errors between what was emitted and what was received, as illustrated in Fig.~\ref{fig:limitations}.a, for $N=16$ as an example, with the small scale of the deviation shown in the enlarged inset. A statistical analysis of the experimental noise, shown graphically in Fig.~\ref{fig:limitations}.b, reveals that $\delta \mathcal{C}_\text{still} \approx 3.7\times10^{-3}$ in still air (no turbulence), which we take to represent the inherent noise of the detection system.  When instantaneous turbulent channel conditions are introduced we observe only a slight increase in noise to $\tilde{\delta \mathcal{C}}_\text{turb} \approx 4.6\times10^{-3}$, which we attribute to the signal-to-noise limit of our detector, i.e., defocusing aberrations in turbulence leading to a signal below the noise threshold of the detector (see Supplementary Information). It is notable that a spread in the received $\mathcal{C}$ due to detector noise is akin to the spectral spread observed in OAM due to channel aberrations.  We confirm too that the noise is not mode specific, as shown by the small difference in variance with and without turbulence in Fig.~\ref{fig:limitations}.c.  This validates the claim that the minimum subdivision of the encoding space, or maximum number of modes (or bits), is limited primarily by detector noise and not channel conditions, contrary to traditional MDM schemes. 

To probe the potential of this approach we consider how a given detector noise limit ($\delta\mathcal{C}$) and subdivision choice ($N$) affect the resulting bit-error-rate (BER), using our turbulence channel as an example. The BER is the ratio of incorrectly received bits to total transmitted bits and represents how crosstalk in the scheme affects the information transfer. The results of this are illustrated graphically in Fig.~\ref{fig:limitations}.d.  In order to quantify the relationship between the cross-talk inducing $\delta\mathcal{C}$ and the system fidelity we can analytically inspect the overlap of neighbouring basis elements to reveal the BER for a given choice of basis (i.e., $N$) and $\delta\mathcal{C}$

\begin{equation}
    \text{BER}(N,\delta\mathcal{C}) = \bar{E}(N)\frac{\delta\mathcal{C}}{2}\sqrt{\frac{\pi}{2}}e^{-\frac{N^{-2}}{2\delta\mathcal{C}}}\text{erf}\left(\sqrt{\frac{2}{\delta\mathcal{C}}}\right)\,,
\end{equation}

where $\text{erf}()$ represents the error function and $\bar{E}(N)$ is the mean bit-error per erroneously binned measurement for a given basis choice (see Supplementary Information). If we inspect the BER for different channel conditions and basis choices, as shown in Fig.~\ref{fig:limitations}.d, we notice how the reducing $\delta \mathcal{C}$ allows for the use of denser choice of basis while maintaining a low BER. The white curve in Fig.~\ref{fig:limitations}.d indicates the special cases of $N \delta \mathcal{C} = 1$, with the associated line acting as a limit for the basis size under a given channel condition. It is however notable that this overlap is significantly lower than the modal overlap experienced by MDM systems, and that with a suitable detector acceptable BERs on the order of $10^{-9}$ are plausible. These results indicate that the effectiveness of the technique for high dimensional information transfer places the burden only on the signal-to-noise ratio of the detector system. The notably poor diffraction efficiency offered by DMDs \cite{scholes2019structured} and the limited dynamic range of CCDs means that our particular system suffers from a lower signal-to-noise ratio (SNR) than alternatives, such as the sensitive photodiodes used in free-space communication links. The white dashed lines in Fig.~\ref{fig:limitations}.d indicate the mean measured $\delta\mathcal{C}$ for our system with and without turbulence. The associated $\text{BER}(N=8)\approx2\times10^{-3}$ also agrees well with the fidelities reported in Fig.~\ref{fig:results}.c. 


\begin{figure*}[t]
    \centering
    \includegraphics{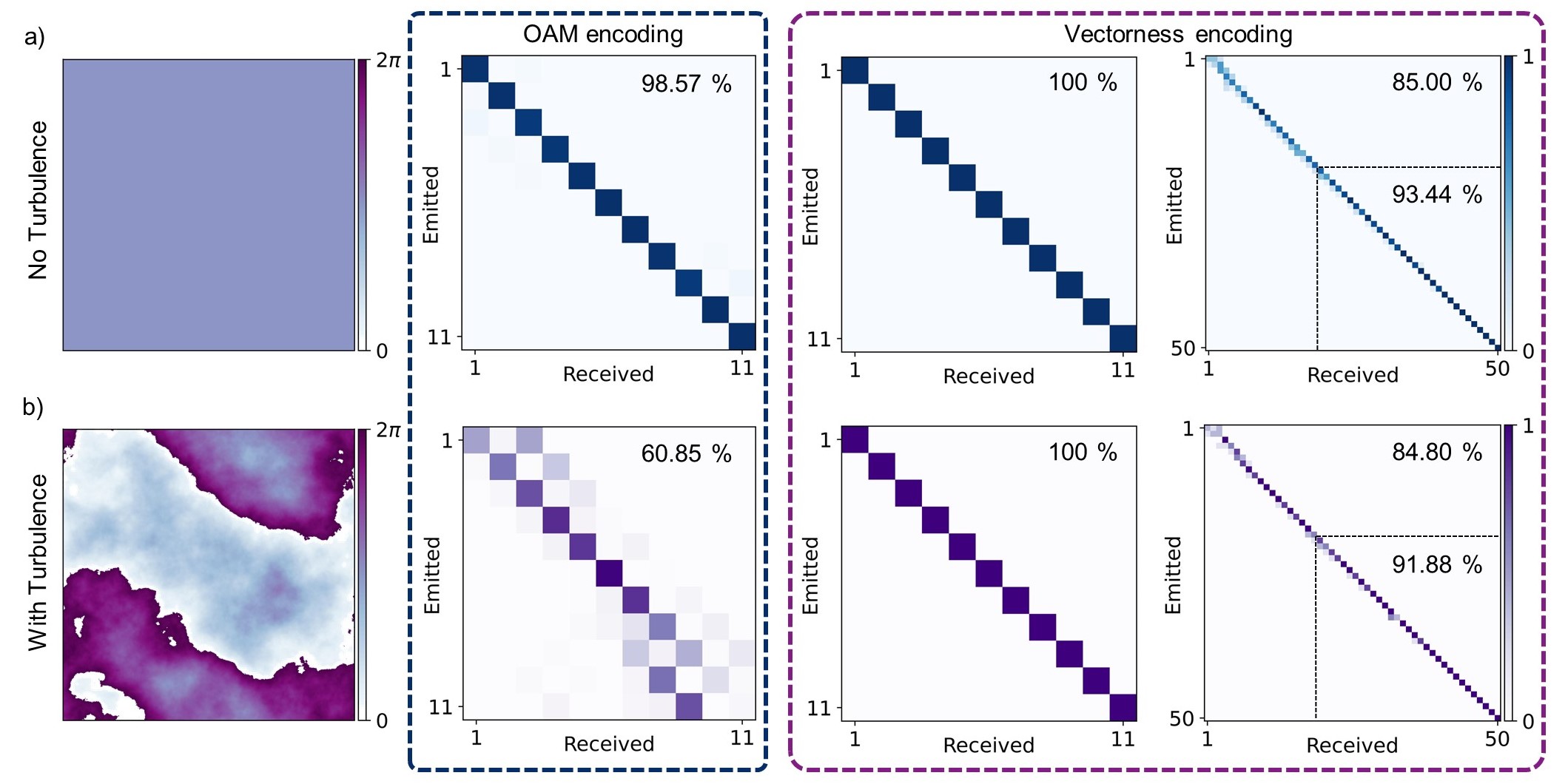}
    \caption{\textbf{Channel invariance comparison to OAM encoding.} Crosstalk matrices through a) ideal and b) aberrated channels using both OAM (with $N=11$) and vectorness (with $N=11$ and $N=50$) encoding schemes. The turbulence phase screen used to aberrate the beams is shown in the left panel of b). A truncated 32 mode subset of the $N=50$ basis is highlighted showing improved fidelity.}
    \label{fig:LargeN}
\end{figure*}

\vspace{0.5cm}
\noindent\textbf{Comparative performance.} The highlight of the proposed vectorness based encoding scheme is that its invariance to the channel conditions makes it a more robust choice in comparison to direct modal encoding schemes.  To illustrate this point we will use the OAM basis as a comparative example. Additionally we changed the CCD detector from a 58.81 dB to a 60.62 dB dynamic range model to further highlight the role of detector choice and to allow a larger basis to be explored (more subdivisions). The setup of Fig.~\ref{fig:setup} (using DMD$_2$) allowed us to generate and perform spatial mode projections on scalar Laguerre-Gaussian beams ($\mathrm{LG}_p^l$) of azimuthal order $l$ and radial order $p$. The results in Fig.~\ref{fig:LargeN}.a show the baseline case with no turbulence for direct OAM encoding and indirect OAM encoding using the vectorness.  In the direct OAM case, we utilized an 11 mode basis spanning the range of $l\in[-5,5]$ with $p=0$. Here the size of the OAM basis was constrained by the system aperture, as increasing mode order results in ever larger beams. Our modal vectorness scheme can be executed with low order modes for the same size across all bit-levels, and so does not suffer from this limitation. The detector noise limit in the vectorness scheme allowed us to probe a $N=50$ basis, far larger than OAM encoding approaches which typically range from $N=4$ to $N=34$ \cite{wang2012terabit,zhu2021compensation}. To the best of our knowledge $N=50$ is the largest number of vectorial modes used in a communication link experiment. The fidelities in Fig.~\ref{fig:LargeN}.a are indicative of experimental limitations (detector noise) and can be taken as the baseline values.

Since we are interested in the effect of the channel, we replaced our dynamic channel aberrations with a static turbulence phase screen with a fixed Fried parameter of $r_0=1.5$ mm as displayed in the left panel of Fig.~\ref{fig:LargeN}.b (see Methods for the generation technique). The static screen allowed for a controlled comparison of how the crosstalk in each of the schemes changes when the channel perturbations are introduced.  The 37.72 \% decrease in fidelity using OAM encoding is in stark contrast to the 0 \% change observed using vectorness encoding at the same basis size ($N=11$).  When pushing the dimensionality of the new encoding scheme further, the channel induced crosstalk only increases marginally, in the order of 1 \%.

There are two important observations one can make from the vectorness data in Fig \ref{fig:LargeN}.  Firstly, the baseline (no turbulence) indicates how the detector choice limits the maximum subdivision number by its noise, setting a limit on how small $\delta C$ can be.  For our detector at $N=50$ the baseline fidelity is 85 \%, but can be improved to 93 \% by using a smaller number of modes ($N<N_\mathrm{max}$) in the available space - while also maintaining the modal density.  With better detector technology the maximum number of modes can be significant, or the impact of noise made small by using a smaller encoding alphabet.  Secondly, a comparison of the baseline (no turbulence) to the baseline with channel noise (with turbulence) makes crystal clear that the channel noise is negligible in the new scheme, validating our central claim.  In no other modal approach is the channel noise so inconsequential, here contributing in the $ \approx 1$ \% range. 


\section*{DISCUSSION AND CONCLUSION}

The proposed scheme and the results of our demonstration reveal its potential for multi-bit encoding using a new modal version of amplitude modulation, enabling a $d$-fold increase in information density using existing amplitude modulation technology for superior data transmission rates. Unlike MDM systems, our source of noise is primarily detector based, while being almost completely invariant to channel aberrations, negating the need for adaptive error-correction, which we showed using atmospheric turbulence as an extreme example.  This is illustrated by using 50 spatial modes in a turbulent channel with a cross-talk that is comparable to the no-turbulence case, differing by less then 2 \%. For comparison, our crosstalk observed between 11 OAM modes was 37.72 \% but zero for the vectorness approach, while other studies with 4 orthogonal vector OAM modes have reported $\approx20$\% crosstalk even in weak turbulence conditions \cite{cox2016resilience}, while crosstalk between 7 neighbouring scalar OAM modes reached $13.2$\% even with corrective measures \cite{zhou2021multiprobe}. Our observed BER using 8 modes, on the order of $2\times10^{-3}$, is comparable to that achieved in error-corrected MDM systems \cite{Wang2012}, even though we have deployed only a proof-of-principle version of the experiment with rudimentary detectors.  Our direct comparison to OAM encoding revealed the superior channel invariance while also demonstrating how a larger basis (e.g., up to 50) can be achieved with the same system aperture which constrained the OAM basis to 11 modes - this is since the increase of basis size in OAM encoding requires larger mode orders which require larger apertures, while our approach can scale the number of modes up to the noise floor of the detector. 

We point out that while the particular vector beams we used were composed of OAM endowed $\mathrm{LG}_p^l$ modes, the core feature of our scheme is the utilization of vector beams (our modal basis) in a manner that does not require the spatial modes nor orthogonal polarisation components to be detected or recognised. Instead, only their vectorness is detected, an invariant quantity that is found by an \textit{integrated} modal signal.  It is a modal approach without the penalty of detecting modes.  In this regard, while our proof-of-principle experiment used a CCD camera for detection, fast and sensitive photo-detectors are all that is required for a real-world implementation. Because our channel-invariant basis is derived from spatial modes without actually detecting them and is independent of the polarisation basis that they were prepared in, it brings with it some significant benefits over traditional MDM schemes that use the modes themselves as the basis. For instance, (i) system misalignment is mitigated since the integrated detection (a power) is spatially invariant, (ii) the modes can be selected with low order to reduce divergence \cite{willner2021perspective} (since more information does not require higher mode numbers), (iii) like amplitude modulation, our scheme does not prohibit other enhancement schemes such as wavelength division multiplexing (WDM) to further improve information density, including even MDM if modal correction is applied, and further, (iv) unlike alternative vectorial techniques \cite{Milione2015d,zhu2021compensation}, the nature of our detection scheme is over-complete even in the polarisation basis, and therefore means that the sender and receiver do not have to agree on the measurement basis, allowing the scheme to work even in polarization scrambling media such as stressed optical fibre. These benefits are provided by the vectorness, a physical quantity which displays resilience to transformations which generally hinder modal communication systems. The trade-off lies in the scaling of our scheme, in order to increase $d$ by $1$, $N$ has to double. This places the burden of performance on the power and amplitude modulation resolution of the emitter and the SNR of the receiver - all factors which already receive considerable attention in free space optical communication systems.

To conclude, we have presented how the vectorness of vectorial light can be used as a new modal version of amplitude modulation, exploiting spatial modes in a manner that makes them channel-invariant, with the number of modes used limited primarily by the sensitivity of the detectors used. We have demonstrated high fidelity, correction-free, multi-bit information transfer to verify our technique, even through dynamic turbulence, an extreme example of a communications channel. Our approach can be extended to other channels too, such as optical fibre and under-water, since the invariance property will hold in all such channels. We believe this approach will open up a new avenue for high-bandwidth optical communication, with the immediate benefits of MDM but without the modal cross-talk challenges. The vectorness presents a new scheme for information encoding, which if coupled with other communication techniques such as WDM and SDM, can be used to push the boundaries of optical communication.

\newpage
\clearpage

\section*{METHODS}

\noindent \textbf{Experimental implementation.} In order to demonstrate the effective high dimensional information transfer using concurrence as the communication basis, we utilized the setup shown in Fig.~\ref{fig:concept}.b. A Gaussian beam produced by a HeNe laser (wavelength 633 nm) was expanded and collimated using lenses L$_1$ and L$_2$ respectively. The plane of polarization was converted to $45^\circ$ using a half-wave plate (HWP$_1$) before passing through a Wollaston prism (WP) which separated the horizontally and vertically polarized components of the expanded beam at an angle of $\approx1^\circ$. The plane at the WP was imaged onto the screen of a digital micro-mirror device (DMD$_1$) using a $4f_1$ imaging system. DMD$_1$ was addressed using two multiplexed binary holograms of the form
\begin{align}
    H_{A/B}(\Vec{r}) &= \frac{1}{2} + \frac{1}{2}\text{sign}[\cos(\varphi_{A/B}(\Vec{r}_T) + 2\pi(\Vec{g}^{A/B}_T\cdot\Vec{r}_T))\nonumber\\ &- \cos(A_{A/B}(\Vec{r}_T))],
    \label{eqn:dmd_hol}
\end{align}

facilitating the modulation of the complex field $U_{A/B}=A_{A/B}e^{i\varphi_{A/B}}$, where $\Vec{g}^{A/B}_T=(g_x^{A/B},g_y^{A/B})$ are grating frequencies \cite{lee1979binary}. In order to control the relative amplitudes of the resulting beams, complimentary weighted random matrices were multiplied to $H_{A/B}$ according to the method outlined in Ref. \cite{hu2021high}. Examples of the multiplexed gratings corresponding the different concurrence values are shown in Fig. \ref{fig:concept}.c. By selecting $\vec
{g}^{A/B}$ appropriately, the $+1$ diffraction orders of independently modulated, orthogonally polarized components were spatially overlapped - creating our vector beam $|\Psi\rangle$ \cite{rosales2020polarisation}. For our case $|\psi_{\pm}\rangle \equiv LG_0^{\pm1}$ were chosen, where $LG_p^l$ is the Laguerre-Gaussian mode with azimuthal(radial) index $l(p)$. The combined diffraction order was isolated using an aperture placed at the focal plane of a $4f_2$ imaging system - which imaged $|\Psi\rangle$ onto a second DMD$_2$. DMD$_2$ was addressed by simple binary diffraction gratings which were multiplexed in order to produce four copies of $|\Psi\rangle$, these copies were allowed to propagate into the far field through a 200 mm length of air - heated by a plate at 185$^\circ$C. Additionally 3 of the four paths were each passed through a HWP$_{2,3}$ (fast axes at 45 and 22.5$^\circ$, respectively) or a quater-wave plate (QWP - fast axis at 45$^\circ$). Lenses L$_3$ and L$_4$ were used to demagnify the four beams onto a CCD camera, while filtering using a linear polarizer (LP - transmission axis at 0$^\circ$). The intensities projected onto the CCD correspond to the horizontal, vertical, diagonal and right-circular polarized components - which allowed for the determination of $\mathcal{C}$ \cite{singh2020digital}. Simulated examples of the far field Stokes intensities showing the region of integration (dashed circles) are given in Fig. \ref{fig:concept}.d.

\textbf{Concurrence measurement.} The Stokes measurements were used to calculate $C$ according to \cite{selyem2019basis}
 
 \begin{align}
     &\mathcal{C} = \nonumber\\&\sqrt{1-\frac{(P_H - P_V)^2+(2P_D-P_H-P_V)^2+(2P_R-P_H-P_V)^2}{(P_H+P_V)^2}}\,,
 \end{align}
 
 where $P_i=\int\int I_idr_T$ represent the powers obtained from the transversely integrated Stokes intensities, the dashed circle in Fig. \ref{fig:setup}.c indicate the necessary region of integration. \\
 
 \textbf{Static turbulence screens.} In order to probe the performance of our encoding scheme along with alternatives under controllable conditions we implemented a static turbulence phase screen, $\Phi(\vec{r}_T)$ which was used to modulate the phase of our beams during their generation (i.e. on DMD$_1$). The screen was generated using a fast-Fourier transform technique according to the following expression
 
 \begin{equation}
     \Phi(\vec{r}_T) = \Re\left(\mathcal{F}^{-1}(M_{rand}\sqrt{\theta})\right)\,,
 \end{equation}
 
where $\Re$ is the real-part, $M_{rand}$ is a complex Gaussian random matrix centered at 0 with unit deviation and $\theta$ is the Kolmogorov-Weiner spectrum - conventionally determined (in pixel coordinates $(i,j)$) by the following expression
 
 \begin{equation}
     \theta(i,j) = 0.023\left(\frac{2D}{r_0}\right)^{\frac{5}{3}}(i^2+j^2)^{-\frac{11}{3}}\,,
     \label{eqn:turb_spec}
 \end{equation}
 
 where $D$ is the hard aperture of the screen and $r_0$ is the Fried parameter (diameter over which the root-mean-squared wavefront aberration is $1$ radian) \cite{Fried1966}. It has been found that Eqn. \ref{eqn:turb_spec} does not accurately model the statistics in the low spatial frequency region (which is notably dominant in the Kolmogorov regime). In order to accurately model this region of the spectrum we employed additional subharmonic sampling according to the method outlined in Ref. \cite{lane1992simulation} - when calculating our spectrum. In our experiment we set $r_0=1.5$ mm.\\
 
\textbf{Modal projection measurements.} To facilitate a comparison to OAM based encoding we generated scalar $U_{emitted}=LG_p^l$ modes by setting $a=0$ on DMD$_1$. For simplicity only the OAM associated azimuthal index $l$ was varied on the interval $[-5,5]$ to form the communication basis - while setting $p=0$. To determine crosstalk between basis elements we utilized correlation filters encoded on the second DMD$_2$ along with the far-field propagation to perform inner product measurements with each of the basis elements. The correlation filters were implemented according to Eqn \ref{eqn:dmd_hol} with $U_A=(LG_0^l)^*$ - resulting in the following intensity profile, to be captured in the far-field by the CCD,
 
 \begin{equation}
     I_{FF} = \left| \int U_{emitted}(LG_0^l)^*e^{i2\pi\vec{k}_T\cdot\vec{r}_T} d\vec{r}_T\right|^2\,,
 \end{equation}
 
where it is clear that measuring the on axis intensity in the far-field (i.e where $\vec{k}_T=0$) results in the absolute-squared inner product between the emitted field and the given basis element. This procedure is referred to as modal decomposition and a thorough tutorial is available in Ref. \cite{pinnell2020modal}.

\newpage

\section*{Author contributions}
The experiment was performed by K.S. and I.N. assisted with the theory.  A.D. and W.B. assisted with editing the manuscript and data analysis.  All authors contributed to writing the manuscript. A.F. conceived of the idea and supervised the project.

\section*{Competing Interests}
The authors declare no competing interests.

\section*{Data availability}
The data that supports the plots within this paper and other findings of this study are available from the corresponding author upon reasonable request.

\newpage



\newpage \clearpage

\setcounter{equation}{0}
\setcounter{figure}{0}   
\renewcommand{\theequation}{S.\arabic{equation}}
\renewcommand{\thefigure}{S.\arabic{figure}}

\onecolumngrid
\begin{center}
\textbf{Supplementary information: A robust basis for multi-bit optical communication with vectorial light}
\end{center}

\vspace{0.5 cm}

\twocolumngrid

\section{Theory}

\subsection{Concept and general approach}

For the benefit of the reader, we will first introduce a general overview of our approach, whereby the core concepts will be subsequently unpacked in further theoretical detail. Our technique makes use of a family of structured light beams \cite{forbes2021structured} with coupled transverse spatial and polarization components, called vector beams, that can be expressed as the linear superposition state
\begin{equation}
|\Psi\rangle = a_{1}|\psi_1 \rangle |H\rangle + a_2|\psi_2\rangle|V\rangle \,,
\end{equation}
\noindent consisting of the spatial modes (not necessarily orthogonal), $|\psi_{1,2}\rangle$, that are coupled to the linear horizontal ($ |H\rangle$) and vertical ($|V\rangle$) polarization states, respectively. The coefficients are normalised so that $|a_{1}|^2+|a_{2}|^2 =1$. Such beams can have independent (factorizable) polarization and spatial components, and can therefore be written as a separable product of polarization and spatial degrees of freedom, often called a scalar beam. Scalar beams are associated with uniform polarization field vectors. Contrarily, the polarization and spatial components can be written as a non-separable product, analogous to quantum entangled two photon states. In this case, the resulting beams have non-uniform polarization field vectors and are called vector beams. In general, the beams can have varying levels of inhomogeneity in the polarization field vectors.

The degree of inhomogeneity of the field vectors can be directly related to the degree of non-separbility between the polarization and spatial components of the field. A measure that accurately quantifies this, is the vector quality factor (VQF) or vectorness \cite{ndagano2016beam} which is a measure that is derived from concurrence \cite{wootters1982single}, commonly used to quantify how entangled (non-separable) independent subsystems are. It ranges from $\mathcal{C}=0$ for completely scalar to $\mathcal{C}=1$ for completely non-separable vector fields.

For channels (or propagation media) that are completely unitary, meaning that only a change of basis in field components, i.e. $| \psi_{1,2} \rangle$, is performed, local rotations of the field vectors are however, still observed upon propagation, although the field retains the same degree of non-separability \cite{nape2018self}. For example, a one-sided channel that only perturbs the spatial components evolves the initial field to the final state
\begin{equation}
|\tilde{\Psi}\rangle = a_1|\phi_1\rangle |H\rangle + a_2|\phi_2\rangle|V\rangle \,,
\label{eq:VM}
\end{equation}
\noindent following the mapping $| \psi_{1,2} \rangle \rightarrow | \phi_{1,2} \rangle $. The resulting concurrence, $\tilde{\mathcal{C}}=\mathcal{C}$, of the input and output states are equivalent \cite{nape2018self}.

When $a_2 = \sqrt{1-a_{1}^2}$, and $ | \psi_{1,2} \rangle$ are orthogonal, then the concurrence is a monotonic function of the parameter $a_1$, and can therefore be used as an encoding basis, where a desired concurrence is encoded by controlling the inhomogeneity of the field via $a_1$. For simplicity in the main text we choose $a_1=\sqrt{a}$.  In the sections that follow, we elucidate this concept further and show that we can design a correction-free communication scheme with concurrence as our multi-bit basis. Next, we show how the concurrence can be measured in a basis independent manner using traditional Stokes polarimetry, a crucial step towards demonstrating our scheme.

\subsection{Basis independent non-separability measurement}

The degree of non-separability (equivalently concurrence), between the polarization and spatial components of a classical vector field can be retrieved from the global Stokes parameters (i.e $S_i=\int\Tilde{S}_i(\Vec{r}_T)d^2 r_T$, where $\Tilde{S}_i(\Vec{r}_T)$ are the transversely varying local Stokes parameters) \cite{selyem2019basis}. 
Consider the field in Eq. (\ref{eq:VM}), having spatial components that are not necessarily orthogonal and setting $a_{1} = a_{2}$. We can use the Gram-Schmidt orthonormalization procedure to retrieve orthogonal normalized spatial fields $|\psi_{\pm}\rangle$ given by

\begin{align}
    &|\psi_{\pm}\rangle = \mathcal{A}_{\pm}\left(\frac{|\psi_1\rangle}{\sqrt{|\langle\psi_1|\psi_1\rangle|}} \pm \frac{|\psi_2\rangle}{\sqrt{|\langle\psi_2|\psi_2\rangle|}}\right) \label{eqn:gs_1}\,,\\
    &\mathcal{A}_\pm = \left(2 \pm 2\frac{\langle\psi_1|\psi_2\rangle}{\sqrt{|\langle\psi_1|\psi_1\rangle||\langle\psi_2|\psi_2\rangle|}}\right)^{-\frac{1}{2}}\label{eqn:gs_2}\,.
\end{align}

This allows us to express $|\Psi\rangle$ in terms of orthonormal spatial field and polarization components according to
\begin{equation}
    |\Psi\rangle = a|\psi_+\rangle|H\rangle + b|\psi_-\rangle|H\rangle + c|\psi_+\rangle|V\rangle + d|\psi_-\rangle|V\rangle \,.
    \label{eqn:special}
\end{equation}
Under the assumptions of completely polarized (pure) states and expressing in terms of orthonormal spatial components (i.e. a qubit system), the following convenient expression for the concurrence can be used \cite{wootters1998entanglement}
\begin{align}
    &\mathcal{C} = 2|ad - bc|\label{eqn:c_concurrence},\\
    &\mathcal{C} = 2\sqrt{\langle\psi_1|\psi_1\rangle\langle\psi_2|\psi_2\rangle - |\langle\psi_1|\psi_2\rangle|^2}\label{eqn:f_concurrence}\,.
\end{align}

Equation \ref{eqn:f_concurrence} follows from substitution of Eqs \ref{eqn:gs_1} and \ref{eqn:gs_2} into Eq. \ref{eqn:c_concurrence}. The concurrence is conveniently retrieved in terms of the original unnormalized arbitrary states.

We can now examine the inner products in the position basis (using  $\int  | \Vec{r}_T  \rangle \langle \Vec{r}_T | d^2r_T = \mathbb{I}$) where \cite{toppel2013goos}
\begin{align}
    &|\psi_{1(2)}(\Vec{r}_T)|e^{i\varphi_{1(2)}(\Vec{r}_T)} =  \langle{ \Vec{r}_T |\psi_{1(2)} } \rangle \,,\\
     &\langle\psi_{1(2)}|\psi_{1(2)}\rangle = \int|\psi_{1(2)}(\Vec{r}_T)|^2 d^2r_T \,,\\
    \& &\langle\psi_{1(2)}|\psi_{2(1)}\rangle = \int|\psi_{1(2)}(\Vec{r}_T)||\psi_{2(1)}(\Vec{r}_T)|e^{\pm i\Delta\varphi} d^2r_T \,,
\end{align}
\noindent where $|\psi_{1(2)}(\Vec{r}_T)|$ and $\varphi_{1(2)}(\Vec{r}_T)$ are the respective amplitude and phase of the orthogonally polarized field components in the given transverse plane. While $\Delta\varphi=\varphi_2(\vec{r}_T)-\varphi_1(\vec{r}_T)$. Comparing the above to the following well known expressions for the Stokes parameters
\begin{align}
    &\tilde{S_0}(\Vec{r}_T) = |\psi_{1}(\Vec{r}_T)|^2 + |\psi_{2}(\Vec{r}_T)|^2 \,,\\
    &\tilde{S_1}(\Vec{r}_T) = |\psi_{1}(\Vec{r}_T)|^2 - |\psi_{2}(\Vec{r}_T)|^2 \,,\\
    &\tilde{S_2}(\Vec{r}_T) = |\psi_{1}(\Vec{r}_T)||\psi_{2}(\Vec{r}_T)|\cos(\varphi_2(\Vec{r}_T)-\varphi_1(\Vec{r}_T)) \,,\\
    &\tilde{S_3}(\Vec{r}_T) = |\psi_{1}(\Vec{r}_T)||\psi_{2}(\Vec{r}_T)|\sin(\varphi_2(\Vec{r}_T)-\varphi_1(\Vec{r}_T)) \,,
\end{align}
allows us to reduce the concurrence to an expression dependent only on the global Stokes parameters
\begin{equation}
    \mathcal{C} = \sqrt{1 - \frac{S_1^2 + S_2^2 + S_3^2}{S_0^2}}\, ,
\end{equation}
where $S^2_i = \int |\tilde{S_i}(\Vec{r}_T)|^2 d^2r_T$, for each stokes parameter. Conventionally the Stokes parameters are determined by the overcomplete set of 6 intensity measurements, we choose to use the minimum number of four measurements of the horizontal ($I_H$), vertical ($I_V$), diagonal ($I_D$) and right-circular ($I_R$) polarization intensity projections to calculate the Stokes parameters as follows

\begin{align}
&\tilde{S_0} = I_H + I_V\,, \\
&\tilde{S_1} = I_H - I_V\,, \\
&\tilde{S_2} = 2I_H + \tilde{S_0}\,, \\
&\tilde{S_3} = 2I_D + \tilde{S_0}\,. 
\end{align}

Furthermore, since the global integration is linear the intensity projections can be directly integrated - leading to the direct determination of the concurrence according to

 \begin{align}
     &\mathcal{C} = \nonumber\\&\sqrt{1-\frac{(P_H - P_V)^2+(2P_D-P_H-P_V)^2+(2P_R-P_H-P_V)^2}{(P_H+P_V)^2}}\,,
 \end{align}
 
here $P_i=\int\int I_idr_T$ represent the powers obtained from the transversely integrated Stokes intensities.

\subsection{Propagation through turbulent media}

Now let us consider the effect of atmospheric turbulence, where spatially varying air densities due to both pressure and temperature variations induce a spatially varying refractive index according to the Gladstone-Dale law. We can express the total phase change $\Phi(\vec{r}_T)$ through a width of turbulent medium using the thin screen approximation \cite{herman1990method}. Therefore the concurrence through the channel (including propagation into the far field) is given by 
\begin{align}
    &\tilde{\mathcal{C}}=2\sqrt{\langle\phi_1|\phi_1\rangle\langle\phi_2|\phi_2\rangle - |\langle\phi_1|\phi_2\rangle|^2},\\
    &|\phi_{1,2}\rangle=\mathcal{F}|\tilde{\psi}_{1/2}\rangle\,;\,\langle\phi_{1(2)}| = \langle\tilde{\psi}_{1(2)}|\mathcal{F}^\dagger,\\
    & \langle \Vec{r}_T | \tilde{\psi}_{1,2}\rangle=|\psi_{1,2}(\vec{r}_T)|e^{i(\varphi_{1,2}(\vec{r}_T) + \Phi(\vec{r}_T))}.
\end{align}
\noindent Here the states $| \tilde{\psi}_{1,2} \rangle$ and $| \phi_{1,2} \rangle$ correspond to the aberrated and subsequently propagated optical fields, respectively. The symbol, $\mathcal{F}$, denotes the Fourier transform operator used here to map the optical modes to the far-field. Since $\mathcal{F}$ is a unitary transformation (i.e. $\mathcal{F}^\dagger\mathcal{F}=\mathcal{F}\mathcal{F}^\dagger\mathbb{I}$), it preserves their inner products such that 
\begin{equation}
    \tilde{\mathcal{C}}=2\sqrt{\langle\tilde{\psi}_1|\tilde{\psi}_1\rangle\langle\tilde{\psi}_2|\tilde{\psi}_2\rangle - |\langle\tilde{\psi}_1|\tilde{\psi}_2 \rangle|^2},
\end{equation}
where we get
\begin{align}
    \langle\tilde{\psi}_{1,2}|\tilde{\psi}_{1,2}\rangle&=\int\left||\psi_{1,2}(\vec{r}_T)|e^{i(\varphi_{1,2}(\vec{r}_T) + \Phi(\vec{r}_T))}\right|^2d^2 r_T\nonumber\\&=\langle\psi_{1,2}|\psi_{1,2}\rangle, \\
    \langle\tilde{\psi}_{1,2}|\tilde{\psi}_{2,1}\rangle&=\int|\psi_{1,2}(\vec{r}_T)|^2e^{\pm (i\Delta\varphi + \Phi(\vec{r}_T)-\Phi(\vec{r}_T))}d^2 r_T \nonumber\\&=\langle\psi_{1,2}|\psi_{2,1}\rangle,
\end{align}
therefore $\tilde{\mathcal{C}}=\mathcal{C}$. This result represents a key advantage presented by the spatially independent concurrence as a basis for free-space communication - making it invariant to atmospheric turbulence. An extremely noteworthy outcome of this is that any unitary (i.e. inner product preserving) transformation will preserve the concurrence - therefore our concept can be readily extended to thick medium models of atmospheric turbulence, as well as stress induced birefringence in optical fiber.

\subsection{Forming a basis for communication}

Naturally the concurrence is a continuously variable parameter, however in order for us to utilize this property as a basis for information transfer we will need to discretize it. For convenience we will form our vector beam using orthonormal spatial components
\begin{equation}
    |\Psi\rangle = \sqrt{a}|\psi_+\rangle|H\rangle + \sqrt{1-a}|\psi_-\rangle|V\rangle\,, 
\end{equation}
(i.e. a special case of Eq \ref{eqn:special}) therefore by varying the coefficient
$a$ according to 
\begin{equation}
    a = \frac{1}{2}(1\pm\sqrt{1-\mathcal{C}^2}),
\end{equation}


\noindent we can achieve a linear variation of the concurrence (non-separability) $\mathcal{C}\in[0,1]$. This means that by knowing the degree of non-separability between the polarization and spatial components that we wish to encode, $\mathcal{C}$, we can find a unique coefficient $a$ that translates it into the vector field. For example, a scalar vector field corresponding to $a = 0$ encodes $\mathcal{C}=0$, while $a=1/2$ encodes $\mathcal{C} = 1$. Limitations imposed on the system through the detection (i.e. Stokes measurement) apparatus will result in an upper limit to the number of unique concurrence $N_{max}$ values which can be reliably detected. The dimension $d$ of our encoding scheme (i.e. the number of independently transferable bits per beam) is then given by
\begin{equation}
        d=\log_2N\,,
\end{equation}

where $N$ is the chosen number of basis elements to be used.

\section{\label{sec:level2}Mean bit error}

\begin{figure}[ht]
    \centering
    \includegraphics{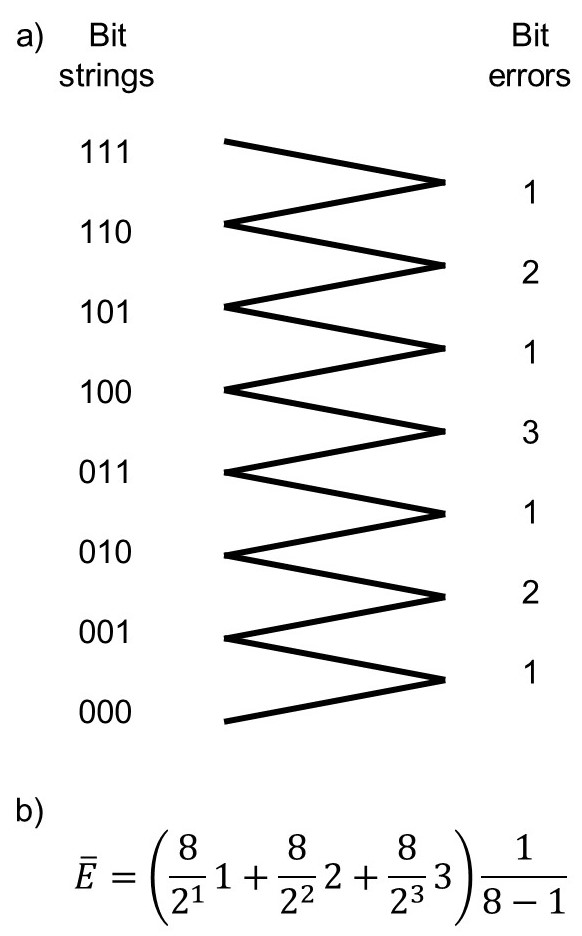}
    \caption{a) Diagram showing the bit errors introduced by erroneous measurements of neighbouring concurrence values. Under the assumption that these errors are distributed homogeneously through the basis the mean can be calculated according to the expression in b).}
    \label{fig:mean_errors}
\end{figure}

The observed crosstalk experienced by the proposed system is generally confined to the neighbouring concurrence values in the discreetized basis. This means that the number of erroneous bits transmitted by an incorrectly binned measurement is dependent on $N$. The mean number of erroneous bits $\bar{E}$ is given by

\begin{equation}
\bar{E}(N) = \sum\limits_{i=1}^d \frac{N}{2^i}\frac{i}{N-1}\,.
\end{equation}

A diagram illustrating the individual and mean errors for a $N=8$ system is shown in Fig. \ref{fig:mean_errors}. The probability $P_{E}$ of an erroneously binned measurement can be approximated using the overlap between neighbouring gaussian distributions given by:

\begin{align}
    P_E(N,\delta\mathcal{C}) &= \int\limits_0^1 e^{-\frac{(\mathcal{C} + (2N)^{-1})^2}{(\delta\mathcal{C})^2}}e^{-\frac{(\mathcal{C} - (2N)^{-1})^2}{(\delta\mathcal{C})^2}}d\mathcal{C}\nonumber\\
    &=\frac{\delta\mathcal{C}}{2}\sqrt{\frac{\pi}{2}}e^{-\frac{N^{-2}}{2\delta\mathcal{C}}}\text{erf}\left(\sqrt{\frac{2}{\delta\mathcal{C}}}\right)\,.
\end{align}

We can then multiply the mean error for a given basis choice by the error probability for that choice in order to determine an effective bit error rate (i.e. $BER(N,\delta\mathcal{C})=\bar{E}(N)P_E(N,\delta\mathcal{C})$).

\section{\label{sec:level1}Impact of Noise}

\begin{figure}
    \centering
    \includegraphics[width=0.7\linewidth]{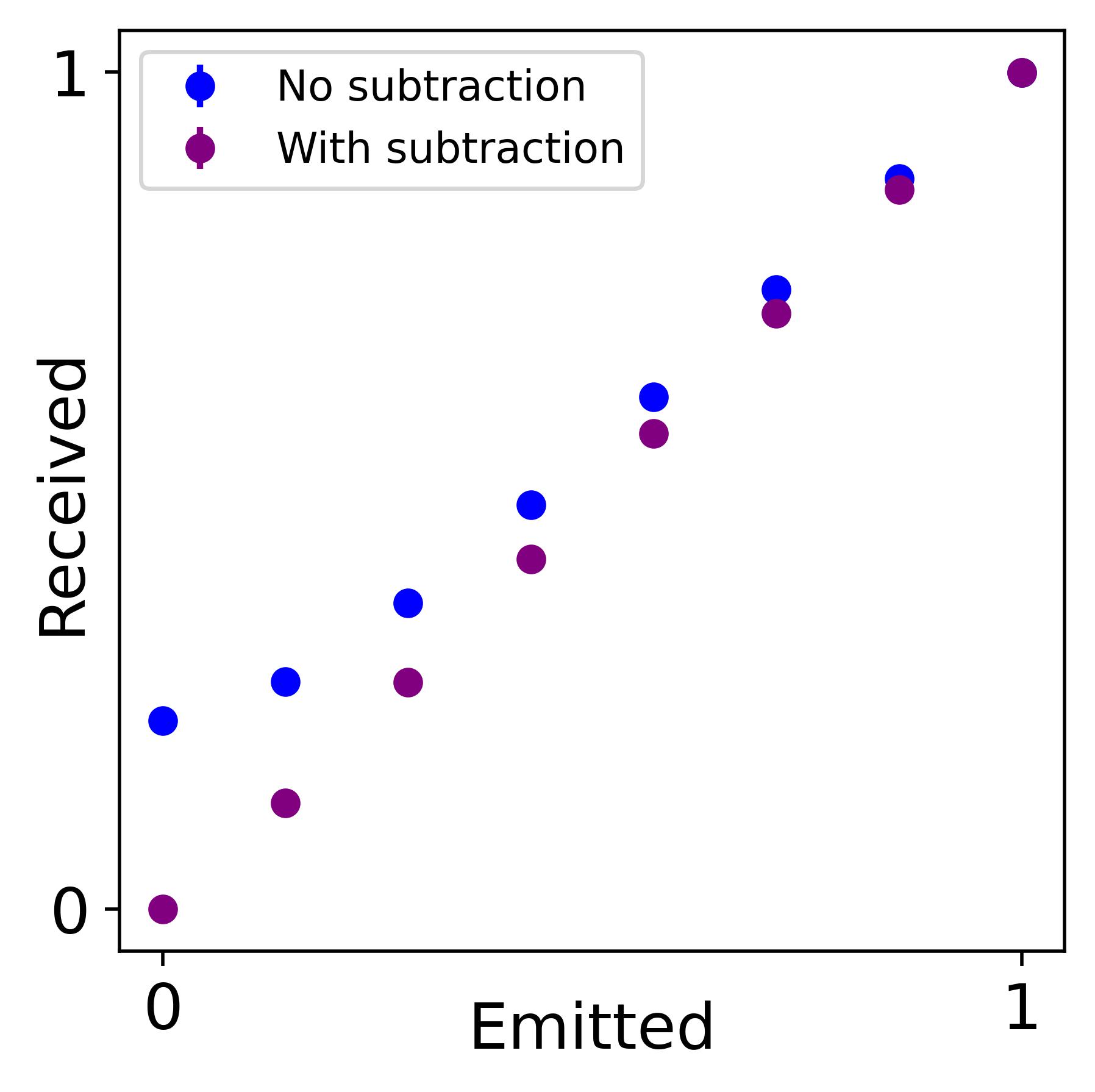}
    \caption{Plot showing the impact of integrated noise subtraction of the received concurrence value. The same Stokes intensity measurements were used to calculate the received concurrence with and without integrated subtraction where the correct (encoded) scaling is achieved with the appropriate subtraction.}
    \label{fig:s1}
\end{figure}

As illustrated in the main text the concurrence is invariant to phase aberrations as they fall under the general umbrella of one-sided (polarization independent) unitary transformations. Noise, however, does impact the concurrence measurement independent of the source (electronic, detector or environmental). The impact of the noise in the region where we expect zeros in the Stokes intensities necessitates the subtraction of a background value from the measurements. The noise subtraction can be done as a single integrated value subtracted from all the powers $P_i$ and allows the measured concurrence to reach zero (see Fig. \ref{fig:s1}). It should be noted that the value of this integrated noise will be dependent on the size of the integration region and the detector shot noise. This subtraction can lead to information loss as parts of the Stokes intensity measurements which fall below the background level are also subtracted - therefore if an integrated Stokes intensity $I_i$ measurement is expected to be below the background level a significant increase in deviation for the encoded vectorness will be observed.

\begin{figure}[h!]
    \centering
    \includegraphics[width=\linewidth]{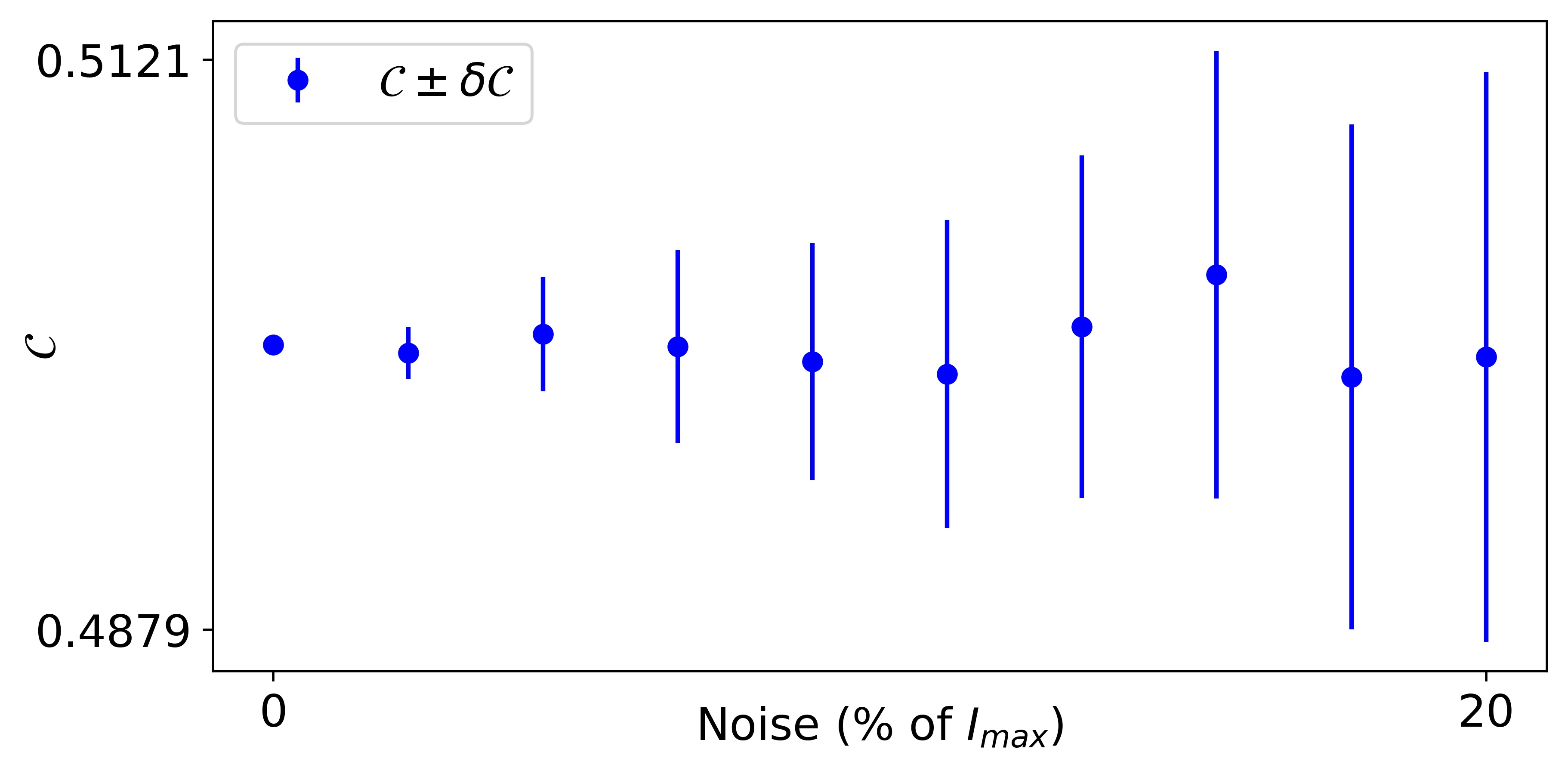}
    \caption{Results of numerical simulations showing the variation of $\delta\mathcal{C}$ (i.e. errorbars) with additive intensity noise (given as a percentage of the maximum Stokes intensity). A vectorness value of $\mathcal{C}=0.5$ was chosen as the target value and a single integrated noise value - appropriate for each noise level - was subtracted.}
    \label{fig:dC_plot} 
\end{figure}

\noindent While the single value noise subtraction provides a convenient way to extract the true encoded vectorness values, spatial and temporal variations in the noise will lead to the over-/under-compensation of the true instantaneous noise conditions. In order to illustrate this we performed numerical simulations where random additive noise was introduced into the far field of a $\mathcal{C}=0.5$ beam. Each Stokes intensity was given a unique noise profile and a single integrated value was subtracted from the integrated powers prior to the vectorness calculation (as was done in the experiments) - 50 simulations were performed at each noise level to acquire the statistics. From the results shown in Fig. \ref{fig:dC_plot}, it is clear that the mean value remains around the target however the deviations $\delta\mathcal{C}$ increase with the noise level as was observed in the experiments.

\bibliography{References}

\providecommand{\noopsort}[1]{}\providecommand{\singleletter}[1]{#1}%
\begin{thebibliography}{79}
\expandafter\ifx\csname natexlab\endcsname\relax\def\natexlab#1{#1}\fi
\expandafter\ifx\csname bibnamefont\endcsname\relax
  \def\bibnamefont#1{#1}\fi
\expandafter\ifx\csname bibfnamefont\endcsname\relax
  \def\bibfnamefont#1{#1}\fi
\expandafter\ifx\csname citenamefont\endcsname\relax
  \def\citenamefont#1{#1}\fi
\expandafter\ifx\csname url\endcsname\relax
  \def\url#1{\texttt{#1}}\fi
\expandafter\ifx\csname urlprefix\endcsname\relax\def\urlprefix{URL }\fi
\providecommand{\bibinfo}[2]{#2}
\providecommand{\eprint}[2][]{\url{#2}}

\bibitem[{\citenamefont{Richardson}(2010)}]{Richardson1}
\bibinfo{author}{\bibfnamefont{D.~J.} \bibnamefont{Richardson}},
  \bibinfo{journal}{Science} \textbf{\bibinfo{volume}{30}},
  \bibinfo{pages}{327} (\bibinfo{year}{2010}).

\bibitem[{\citenamefont{Richardson et~al.}(2013)\citenamefont{Richardson, Fini,
  and Nelson}}]{richardson2013space}
\bibinfo{author}{\bibfnamefont{D.}~\bibnamefont{Richardson}},
  \bibinfo{author}{\bibfnamefont{J.}~\bibnamefont{Fini}}, \bibnamefont{and}
  \bibinfo{author}{\bibfnamefont{L.}~\bibnamefont{Nelson}},
  \bibinfo{journal}{Nature Photonics} \textbf{\bibinfo{volume}{7}},
  \bibinfo{pages}{354} (\bibinfo{year}{2013}).

\bibitem[{\citenamefont{Forbes et~al.}(2021)\citenamefont{Forbes, de~Oliveira,
  and Dennis}}]{forbes2021structured}
\bibinfo{author}{\bibfnamefont{A.}~\bibnamefont{Forbes}},
  \bibinfo{author}{\bibfnamefont{M.}~\bibnamefont{de~Oliveira}},
  \bibnamefont{and} \bibinfo{author}{\bibfnamefont{M.~R.}
  \bibnamefont{Dennis}}, \bibinfo{journal}{Nature Photonics}
  \textbf{\bibinfo{volume}{15}}, \bibinfo{pages}{253} (\bibinfo{year}{2021}).

\bibitem[{\citenamefont{Li et~al.}(2014)\citenamefont{Li, Bai, Zhao, and
  Xia}}]{li2014space}
\bibinfo{author}{\bibfnamefont{G.}~\bibnamefont{Li}},
  \bibinfo{author}{\bibfnamefont{N.}~\bibnamefont{Bai}},
  \bibinfo{author}{\bibfnamefont{N.}~\bibnamefont{Zhao}}, \bibnamefont{and}
  \bibinfo{author}{\bibfnamefont{C.}~\bibnamefont{Xia}},
  \bibinfo{journal}{Advances in Optics and Photonics}
  \textbf{\bibinfo{volume}{6}}, \bibinfo{pages}{413} (\bibinfo{year}{2014}).

\bibitem[{\citenamefont{Berdagu{\'e} and Facq}(1982)}]{Berdague1982A}
\bibinfo{author}{\bibfnamefont{S.}~\bibnamefont{Berdagu{\'e}}}
  \bibnamefont{and} \bibinfo{author}{\bibfnamefont{P.}~\bibnamefont{Facq}},
  \bibinfo{journal}{Appl. Opt.} \textbf{\bibinfo{volume}{21}},
  \bibinfo{pages}{1950} (\bibinfo{year}{1982}).

\bibitem[{\citenamefont{Trichili et~al.}(2019)\citenamefont{Trichili, Park,
  Zghal, Ooi, and Alouini}}]{trichili2019communicating}
\bibinfo{author}{\bibfnamefont{A.}~\bibnamefont{Trichili}},
  \bibinfo{author}{\bibfnamefont{K.-H.} \bibnamefont{Park}},
  \bibinfo{author}{\bibfnamefont{M.}~\bibnamefont{Zghal}},
  \bibinfo{author}{\bibfnamefont{B.~S.} \bibnamefont{Ooi}}, \bibnamefont{and}
  \bibinfo{author}{\bibfnamefont{M.-S.} \bibnamefont{Alouini}},
  \bibinfo{journal}{IEEE Communications Surveys \& Tutorials}
  \textbf{\bibinfo{volume}{21}}, \bibinfo{pages}{3175} (\bibinfo{year}{2019}).

\bibitem[{\citenamefont{Trichili et~al.}(2020)\citenamefont{Trichili, Cox, Ooi,
  and Alouini}}]{trichili2020roadmap}
\bibinfo{author}{\bibfnamefont{A.}~\bibnamefont{Trichili}},
  \bibinfo{author}{\bibfnamefont{M.~A.} \bibnamefont{Cox}},
  \bibinfo{author}{\bibfnamefont{B.~S.} \bibnamefont{Ooi}}, \bibnamefont{and}
  \bibinfo{author}{\bibfnamefont{M.-S.} \bibnamefont{Alouini}},
  \bibinfo{journal}{JOSA B} \textbf{\bibinfo{volume}{37}},
  \bibinfo{pages}{A184} (\bibinfo{year}{2020}).

\bibitem[{\citenamefont{Padgett}(2017)}]{padgett2017orbital}
\bibinfo{author}{\bibfnamefont{M.~J.} \bibnamefont{Padgett}},
  \bibinfo{journal}{Optics Express} \textbf{\bibinfo{volume}{25}},
  \bibinfo{pages}{11265} (\bibinfo{year}{2017}).

\bibitem[{\citenamefont{Willner et~al.}(2015)\citenamefont{Willner, Huang, Yan,
  Ren, Ahmed, Xie, Bao, Li, Cao, Zhao et~al.}}]{Willner2015}
\bibinfo{author}{\bibfnamefont{A.~E.} \bibnamefont{Willner}},
  \bibinfo{author}{\bibfnamefont{H.}~\bibnamefont{Huang}},
  \bibinfo{author}{\bibfnamefont{Y.}~\bibnamefont{Yan}},
  \bibinfo{author}{\bibfnamefont{Y.}~\bibnamefont{Ren}},
  \bibinfo{author}{\bibfnamefont{N.}~\bibnamefont{Ahmed}},
  \bibinfo{author}{\bibfnamefont{G.}~\bibnamefont{Xie}},
  \bibinfo{author}{\bibfnamefont{C.}~\bibnamefont{Bao}},
  \bibinfo{author}{\bibfnamefont{L.}~\bibnamefont{Li}},
  \bibinfo{author}{\bibfnamefont{Y.}~\bibnamefont{Cao}},
  \bibinfo{author}{\bibfnamefont{Z.}~\bibnamefont{Zhao}}, \bibnamefont{et~al.},
  \bibinfo{journal}{Advances in Optics and Photonics}
  \textbf{\bibinfo{volume}{7}}, \bibinfo{pages}{66} (\bibinfo{year}{2015}).

\bibitem[{\citenamefont{Willner et~al.}(2021)\citenamefont{Willner, Pang, Song,
  Zou, and Zhou}}]{willner2021orbital}
\bibinfo{author}{\bibfnamefont{A.~E.} \bibnamefont{Willner}},
  \bibinfo{author}{\bibfnamefont{K.}~\bibnamefont{Pang}},
  \bibinfo{author}{\bibfnamefont{H.}~\bibnamefont{Song}},
  \bibinfo{author}{\bibfnamefont{K.}~\bibnamefont{Zou}}, \bibnamefont{and}
  \bibinfo{author}{\bibfnamefont{H.}~\bibnamefont{Zhou}},
  \bibinfo{journal}{Applied Physics Reviews} \textbf{\bibinfo{volume}{8}},
  \bibinfo{pages}{041312} (\bibinfo{year}{2021}).

\bibitem[{\citenamefont{Wang et~al.}(2014)\citenamefont{Wang, Li, Luo, Liu,
  Zhu, Li, Xie, Yang, Yu, Sun et~al.}}]{wang2014n}
\bibinfo{author}{\bibfnamefont{J.}~\bibnamefont{Wang}},
  \bibinfo{author}{\bibfnamefont{S.}~\bibnamefont{Li}},
  \bibinfo{author}{\bibfnamefont{M.}~\bibnamefont{Luo}},
  \bibinfo{author}{\bibfnamefont{J.}~\bibnamefont{Liu}},
  \bibinfo{author}{\bibfnamefont{L.}~\bibnamefont{Zhu}},
  \bibinfo{author}{\bibfnamefont{C.}~\bibnamefont{Li}},
  \bibinfo{author}{\bibfnamefont{D.}~\bibnamefont{Xie}},
  \bibinfo{author}{\bibfnamefont{Q.}~\bibnamefont{Yang}},
  \bibinfo{author}{\bibfnamefont{S.}~\bibnamefont{Yu}},
  \bibinfo{author}{\bibfnamefont{J.}~\bibnamefont{Sun}}, \bibnamefont{et~al.},
  in \emph{\bibinfo{booktitle}{2014 The European Conference on Optical
  Communication}} (\bibinfo{organization}{IEEE}, \bibinfo{year}{2014}), pp.
  \bibinfo{pages}{1--3}.

\bibitem[{\citenamefont{Zhao et~al.}(2016)\citenamefont{Zhao, Liu, Du, Li, Luo,
  Wang, Zhu, and Wang}}]{zhao2016experimental}
\bibinfo{author}{\bibfnamefont{Y.}~\bibnamefont{Zhao}},
  \bibinfo{author}{\bibfnamefont{J.}~\bibnamefont{Liu}},
  \bibinfo{author}{\bibfnamefont{J.}~\bibnamefont{Du}},
  \bibinfo{author}{\bibfnamefont{S.}~\bibnamefont{Li}},
  \bibinfo{author}{\bibfnamefont{Y.}~\bibnamefont{Luo}},
  \bibinfo{author}{\bibfnamefont{A.}~\bibnamefont{Wang}},
  \bibinfo{author}{\bibfnamefont{L.}~\bibnamefont{Zhu}}, \bibnamefont{and}
  \bibinfo{author}{\bibfnamefont{J.}~\bibnamefont{Wang}}, in
  \emph{\bibinfo{booktitle}{Optical fiber communication conference}}
  (\bibinfo{organization}{Optical Society of America}, \bibinfo{year}{2016}),
  pp. \bibinfo{pages}{Th1H--3}.

\bibitem[{\citenamefont{Bozinivic et~al.}(2013)\citenamefont{Bozinivic, Y.,
  Ren, Tur, Kristensn, Huang, Willner, and Ramachandran}}]{Bozinivic2013}
\bibinfo{author}{\bibfnamefont{N.}~\bibnamefont{Bozinivic}},
  \bibinfo{author}{\bibfnamefont{Y.}~\bibnamefont{Y.}},
  \bibinfo{author}{\bibfnamefont{Y.}~\bibnamefont{Ren}},
  \bibinfo{author}{\bibfnamefont{M.}~\bibnamefont{Tur}},
  \bibinfo{author}{\bibfnamefont{P.}~\bibnamefont{Kristensn}},
  \bibinfo{author}{\bibfnamefont{H.}~\bibnamefont{Huang}},
  \bibinfo{author}{\bibnamefont{Willner}}, \bibnamefont{and}
  \bibinfo{author}{\bibfnamefont{S.}~\bibnamefont{Ramachandran}},
  \bibinfo{journal}{Science} \textbf{\bibinfo{volume}{340}}
  (\bibinfo{year}{2013}).

\bibitem[{\citenamefont{Anguita et~al.}(2008)\citenamefont{Anguita, Neifeld,
  and Vasic}}]{JaimeA.Anguita2008}
\bibinfo{author}{\bibfnamefont{J.~A.} \bibnamefont{Anguita}},
  \bibinfo{author}{\bibfnamefont{M.~A.} \bibnamefont{Neifeld}},
  \bibnamefont{and} \bibinfo{author}{\bibfnamefont{B.~V.} \bibnamefont{Vasic}},
  \bibinfo{journal}{Appl. Opt.} \textbf{\bibinfo{volume}{47}},
  \bibinfo{pages}{2414} (\bibinfo{year}{2008}).

\bibitem[{\citenamefont{Ren et~al.}(2016)\citenamefont{Ren, Wang, Liao, Li,
  Xie, Huang, Zhao, Yan, Ahmed, Willner et~al.}}]{ren2016experimental}
\bibinfo{author}{\bibfnamefont{Y.}~\bibnamefont{Ren}},
  \bibinfo{author}{\bibfnamefont{Z.}~\bibnamefont{Wang}},
  \bibinfo{author}{\bibfnamefont{P.}~\bibnamefont{Liao}},
  \bibinfo{author}{\bibfnamefont{L.}~\bibnamefont{Li}},
  \bibinfo{author}{\bibfnamefont{G.}~\bibnamefont{Xie}},
  \bibinfo{author}{\bibfnamefont{H.}~\bibnamefont{Huang}},
  \bibinfo{author}{\bibfnamefont{Z.}~\bibnamefont{Zhao}},
  \bibinfo{author}{\bibfnamefont{Y.}~\bibnamefont{Yan}},
  \bibinfo{author}{\bibfnamefont{N.}~\bibnamefont{Ahmed}},
  \bibinfo{author}{\bibfnamefont{A.}~\bibnamefont{Willner}},
  \bibnamefont{et~al.}, \bibinfo{journal}{Optics Letters}
  \textbf{\bibinfo{volume}{41}}, \bibinfo{pages}{622} (\bibinfo{year}{2016}).

\bibitem[{\citenamefont{Krenn et~al.}(2014)\citenamefont{Krenn, Fickler, Fink,
  Handsteiner, Malik, Scheidl, Ursin, and Zeilinger}}]{krenn2014communication}
\bibinfo{author}{\bibfnamefont{M.}~\bibnamefont{Krenn}},
  \bibinfo{author}{\bibfnamefont{R.}~\bibnamefont{Fickler}},
  \bibinfo{author}{\bibfnamefont{M.}~\bibnamefont{Fink}},
  \bibinfo{author}{\bibfnamefont{J.}~\bibnamefont{Handsteiner}},
  \bibinfo{author}{\bibfnamefont{M.}~\bibnamefont{Malik}},
  \bibinfo{author}{\bibfnamefont{T.}~\bibnamefont{Scheidl}},
  \bibinfo{author}{\bibfnamefont{R.}~\bibnamefont{Ursin}}, \bibnamefont{and}
  \bibinfo{author}{\bibfnamefont{A.}~\bibnamefont{Zeilinger}},
  \bibinfo{journal}{New Journal of Physics} \textbf{\bibinfo{volume}{16}},
  \bibinfo{pages}{113028} (\bibinfo{year}{2014}).

\bibitem[{\citenamefont{Rodenburg et~al.}(2012)\citenamefont{Rodenburg, Lavery,
  Malik, O'Sullivan, Mirhosseini, Robertson, Padgett, and
  Boyd}}]{rodenburg2012a}
\bibinfo{author}{\bibfnamefont{B.}~\bibnamefont{Rodenburg}},
  \bibinfo{author}{\bibfnamefont{M.~P.~J.} \bibnamefont{Lavery}},
  \bibinfo{author}{\bibfnamefont{M.}~\bibnamefont{Malik}},
  \bibinfo{author}{\bibfnamefont{M.~N.} \bibnamefont{O'Sullivan}},
  \bibinfo{author}{\bibfnamefont{M.}~\bibnamefont{Mirhosseini}},
  \bibinfo{author}{\bibfnamefont{D.~J.} \bibnamefont{Robertson}},
  \bibinfo{author}{\bibfnamefont{M.}~\bibnamefont{Padgett}}, \bibnamefont{and}
  \bibinfo{author}{\bibfnamefont{R.~W.} \bibnamefont{Boyd}},
  \bibinfo{journal}{Optics Letters} \textbf{\bibinfo{volume}{37}},
  \bibinfo{pages}{3735} (\bibinfo{year}{2012}).

\bibitem[{\citenamefont{Zhang et~al.}(2020)\citenamefont{Zhang, Shen, Lan, and
  Tang}}]{zhang2020mode}
\bibinfo{author}{\bibfnamefont{L.}~\bibnamefont{Zhang}},
  \bibinfo{author}{\bibfnamefont{F.}~\bibnamefont{Shen}},
  \bibinfo{author}{\bibfnamefont{B.}~\bibnamefont{Lan}}, \bibnamefont{and}
  \bibinfo{author}{\bibfnamefont{A.}~\bibnamefont{Tang}},
  \bibinfo{journal}{Journal of Optics} \textbf{\bibinfo{volume}{22}},
  \bibinfo{pages}{075607} (\bibinfo{year}{2020}).

\bibitem[{\citenamefont{Malik et~al.}(2012)\citenamefont{Malik, O’Sullivan,
  Rodenburg, Mirhosseini, Leach, Lavery, Padgett, and
  Boyd}}]{malik2012influence}
\bibinfo{author}{\bibfnamefont{M.}~\bibnamefont{Malik}},
  \bibinfo{author}{\bibfnamefont{M.}~\bibnamefont{O’Sullivan}},
  \bibinfo{author}{\bibfnamefont{B.}~\bibnamefont{Rodenburg}},
  \bibinfo{author}{\bibfnamefont{M.}~\bibnamefont{Mirhosseini}},
  \bibinfo{author}{\bibfnamefont{J.}~\bibnamefont{Leach}},
  \bibinfo{author}{\bibfnamefont{M.~P.} \bibnamefont{Lavery}},
  \bibinfo{author}{\bibfnamefont{M.~J.} \bibnamefont{Padgett}},
  \bibnamefont{and} \bibinfo{author}{\bibfnamefont{R.~W.} \bibnamefont{Boyd}},
  \bibinfo{journal}{Optics express} \textbf{\bibinfo{volume}{20}},
  \bibinfo{pages}{13195} (\bibinfo{year}{2012}).

\bibitem[{\citenamefont{Chen et~al.}(2016)\citenamefont{Chen, Yang, Tong, and
  Lou}}]{chen2016changes}
\bibinfo{author}{\bibfnamefont{C.}~\bibnamefont{Chen}},
  \bibinfo{author}{\bibfnamefont{H.}~\bibnamefont{Yang}},
  \bibinfo{author}{\bibfnamefont{S.}~\bibnamefont{Tong}}, \bibnamefont{and}
  \bibinfo{author}{\bibfnamefont{Y.}~\bibnamefont{Lou}},
  \bibinfo{journal}{Optics express} \textbf{\bibinfo{volume}{24}},
  \bibinfo{pages}{6959} (\bibinfo{year}{2016}).

\bibitem[{\citenamefont{Tyler and Boyd}(2009)}]{tyler2009influence}
\bibinfo{author}{\bibfnamefont{G.~A.} \bibnamefont{Tyler}} \bibnamefont{and}
  \bibinfo{author}{\bibfnamefont{R.~W.} \bibnamefont{Boyd}},
  \bibinfo{journal}{Optics letters} \textbf{\bibinfo{volume}{34}},
  \bibinfo{pages}{142} (\bibinfo{year}{2009}).

\bibitem[{\citenamefont{Cox et~al.}(2020)\citenamefont{Cox, Mphuthi, Nape,
  Mashaba, Cheng, and Forbes}}]{cox2020structured}
\bibinfo{author}{\bibfnamefont{M.~A.} \bibnamefont{Cox}},
  \bibinfo{author}{\bibfnamefont{N.}~\bibnamefont{Mphuthi}},
  \bibinfo{author}{\bibfnamefont{I.}~\bibnamefont{Nape}},
  \bibinfo{author}{\bibfnamefont{N.}~\bibnamefont{Mashaba}},
  \bibinfo{author}{\bibfnamefont{L.}~\bibnamefont{Cheng}}, \bibnamefont{and}
  \bibinfo{author}{\bibfnamefont{A.}~\bibnamefont{Forbes}},
  \bibinfo{journal}{IEEE Journal of Selected Topics in Quantum Electronics}
  \textbf{\bibinfo{volume}{27}}, \bibinfo{pages}{1} (\bibinfo{year}{2020}).

\bibitem[{\citenamefont{Mphuthi et~al.}(2018)\citenamefont{Mphuthi, Botha, and
  Forbes}}]{mphuthi2018bessel}
\bibinfo{author}{\bibfnamefont{N.}~\bibnamefont{Mphuthi}},
  \bibinfo{author}{\bibfnamefont{R.}~\bibnamefont{Botha}}, \bibnamefont{and}
  \bibinfo{author}{\bibfnamefont{A.}~\bibnamefont{Forbes}},
  \bibinfo{journal}{JOSA A} \textbf{\bibinfo{volume}{35}},
  \bibinfo{pages}{1021} (\bibinfo{year}{2018}).

\bibitem[{\citenamefont{Mphuthi et~al.}(2019)\citenamefont{Mphuthi, Gailele,
  Litvin, Dudley, Botha, and Forbes}}]{mphuthi2019free}
\bibinfo{author}{\bibfnamefont{N.}~\bibnamefont{Mphuthi}},
  \bibinfo{author}{\bibfnamefont{L.}~\bibnamefont{Gailele}},
  \bibinfo{author}{\bibfnamefont{I.}~\bibnamefont{Litvin}},
  \bibinfo{author}{\bibfnamefont{A.}~\bibnamefont{Dudley}},
  \bibinfo{author}{\bibfnamefont{R.}~\bibnamefont{Botha}}, \bibnamefont{and}
  \bibinfo{author}{\bibfnamefont{A.}~\bibnamefont{Forbes}},
  \bibinfo{journal}{Applied Optics} \textbf{\bibinfo{volume}{58}},
  \bibinfo{pages}{4258} (\bibinfo{year}{2019}).

\bibitem[{\citenamefont{Lukin}(2014)}]{lukin2014mean}
\bibinfo{author}{\bibfnamefont{I.~P.} \bibnamefont{Lukin}},
  \bibinfo{journal}{Applied optics} \textbf{\bibinfo{volume}{53}},
  \bibinfo{pages}{3287} (\bibinfo{year}{2014}).

\bibitem[{\citenamefont{Bao-Suan and Ji-Xiong}(2009)}]{bao2009propagation}
\bibinfo{author}{\bibfnamefont{C.}~\bibnamefont{Bao-Suan}} \bibnamefont{and}
  \bibinfo{author}{\bibfnamefont{P.}~\bibnamefont{Ji-Xiong}},
  \bibinfo{journal}{Chinese Physics B} \textbf{\bibinfo{volume}{18}},
  \bibinfo{pages}{1033} (\bibinfo{year}{2009}).

\bibitem[{\citenamefont{Zhu et~al.}(2008)\citenamefont{Zhu, Zhou, Li, Zheng,
  and Tang}}]{zhu2008propagation}
\bibinfo{author}{\bibfnamefont{K.}~\bibnamefont{Zhu}},
  \bibinfo{author}{\bibfnamefont{G.}~\bibnamefont{Zhou}},
  \bibinfo{author}{\bibfnamefont{X.}~\bibnamefont{Li}},
  \bibinfo{author}{\bibfnamefont{X.}~\bibnamefont{Zheng}}, \bibnamefont{and}
  \bibinfo{author}{\bibfnamefont{H.}~\bibnamefont{Tang}},
  \bibinfo{journal}{Optics Express} \textbf{\bibinfo{volume}{16}},
  \bibinfo{pages}{21315} (\bibinfo{year}{2008}).

\bibitem[{\citenamefont{Nelson et~al.}(2014)\citenamefont{Nelson, Palastro,
  Davis, and Sprangle}}]{nelson2014propagation}
\bibinfo{author}{\bibfnamefont{W.}~\bibnamefont{Nelson}},
  \bibinfo{author}{\bibfnamefont{J.}~\bibnamefont{Palastro}},
  \bibinfo{author}{\bibfnamefont{C.}~\bibnamefont{Davis}}, \bibnamefont{and}
  \bibinfo{author}{\bibfnamefont{P.}~\bibnamefont{Sprangle}},
  \bibinfo{journal}{JOSA A} \textbf{\bibinfo{volume}{31}}, \bibinfo{pages}{603}
  (\bibinfo{year}{2014}).

\bibitem[{\citenamefont{Ahmed et~al.}(2016)\citenamefont{Ahmed, Zhao, Li,
  Huang, Lavery, Liao, Yan, Wang, Xie, Ren et~al.}}]{ahmed2016mode}
\bibinfo{author}{\bibfnamefont{N.}~\bibnamefont{Ahmed}},
  \bibinfo{author}{\bibfnamefont{Z.}~\bibnamefont{Zhao}},
  \bibinfo{author}{\bibfnamefont{L.}~\bibnamefont{Li}},
  \bibinfo{author}{\bibfnamefont{H.}~\bibnamefont{Huang}},
  \bibinfo{author}{\bibfnamefont{M.~P.} \bibnamefont{Lavery}},
  \bibinfo{author}{\bibfnamefont{P.}~\bibnamefont{Liao}},
  \bibinfo{author}{\bibfnamefont{Y.}~\bibnamefont{Yan}},
  \bibinfo{author}{\bibfnamefont{Z.}~\bibnamefont{Wang}},
  \bibinfo{author}{\bibfnamefont{G.}~\bibnamefont{Xie}},
  \bibinfo{author}{\bibfnamefont{Y.}~\bibnamefont{Ren}}, \bibnamefont{et~al.},
  \bibinfo{journal}{Scientific reports} \textbf{\bibinfo{volume}{6}},
  \bibinfo{pages}{22082} (\bibinfo{year}{2016}).

\bibitem[{\citenamefont{Cheng et~al.}(2016)\citenamefont{Cheng, Guo, Li, and
  Zhang}}]{cheng2016channel}
\bibinfo{author}{\bibfnamefont{M.}~\bibnamefont{Cheng}},
  \bibinfo{author}{\bibfnamefont{L.}~\bibnamefont{Guo}},
  \bibinfo{author}{\bibfnamefont{J.}~\bibnamefont{Li}}, \bibnamefont{and}
  \bibinfo{author}{\bibfnamefont{Y.}~\bibnamefont{Zhang}},
  \bibinfo{journal}{IEEE Photonics Journal} \textbf{\bibinfo{volume}{8}},
  \bibinfo{pages}{1} (\bibinfo{year}{2016}).

\bibitem[{\citenamefont{Doster and Watnik}(2016)}]{doster2016laguerre}
\bibinfo{author}{\bibfnamefont{T.}~\bibnamefont{Doster}} \bibnamefont{and}
  \bibinfo{author}{\bibfnamefont{A.~T.} \bibnamefont{Watnik}},
  \bibinfo{journal}{Applied Optics} \textbf{\bibinfo{volume}{55}},
  \bibinfo{pages}{10239} (\bibinfo{year}{2016}).

\bibitem[{\citenamefont{Watkins et~al.}(2020)\citenamefont{Watkins, Dai, White,
  Li, Miller, Morgan, and Johnson}}]{watkins2020experimental}
\bibinfo{author}{\bibfnamefont{R.~J.} \bibnamefont{Watkins}},
  \bibinfo{author}{\bibfnamefont{K.}~\bibnamefont{Dai}},
  \bibinfo{author}{\bibfnamefont{G.}~\bibnamefont{White}},
  \bibinfo{author}{\bibfnamefont{W.}~\bibnamefont{Li}},
  \bibinfo{author}{\bibfnamefont{J.~K.} \bibnamefont{Miller}},
  \bibinfo{author}{\bibfnamefont{K.~S.} \bibnamefont{Morgan}},
  \bibnamefont{and} \bibinfo{author}{\bibfnamefont{E.~G.}
  \bibnamefont{Johnson}}, \bibinfo{journal}{Optics express}
  \textbf{\bibinfo{volume}{28}}, \bibinfo{pages}{924} (\bibinfo{year}{2020}).

\bibitem[{\citenamefont{Vetter et~al.}(2019)\citenamefont{Vetter, Steinkopf,
  Bergner, Ornigotti, Nolte, Gross, and Szameit}}]{vetter2019realization}
\bibinfo{author}{\bibfnamefont{C.}~\bibnamefont{Vetter}},
  \bibinfo{author}{\bibfnamefont{R.}~\bibnamefont{Steinkopf}},
  \bibinfo{author}{\bibfnamefont{K.}~\bibnamefont{Bergner}},
  \bibinfo{author}{\bibfnamefont{M.}~\bibnamefont{Ornigotti}},
  \bibinfo{author}{\bibfnamefont{S.}~\bibnamefont{Nolte}},
  \bibinfo{author}{\bibfnamefont{H.}~\bibnamefont{Gross}}, \bibnamefont{and}
  \bibinfo{author}{\bibfnamefont{A.}~\bibnamefont{Szameit}},
  \bibinfo{journal}{Laser \& Photonics Reviews} \textbf{\bibinfo{volume}{13}},
  \bibinfo{pages}{1900103} (\bibinfo{year}{2019}).

\bibitem[{\citenamefont{Yuan et~al.}(2017)\citenamefont{Yuan, Lei, Li, Li, Gao,
  Xie, and Yuan}}]{yuan2017beam}
\bibinfo{author}{\bibfnamefont{Y.}~\bibnamefont{Yuan}},
  \bibinfo{author}{\bibfnamefont{T.}~\bibnamefont{Lei}},
  \bibinfo{author}{\bibfnamefont{Z.}~\bibnamefont{Li}},
  \bibinfo{author}{\bibfnamefont{Y.}~\bibnamefont{Li}},
  \bibinfo{author}{\bibfnamefont{S.}~\bibnamefont{Gao}},
  \bibinfo{author}{\bibfnamefont{Z.}~\bibnamefont{Xie}}, \bibnamefont{and}
  \bibinfo{author}{\bibfnamefont{X.}~\bibnamefont{Yuan}},
  \bibinfo{journal}{Scientific reports} \textbf{\bibinfo{volume}{7}},
  \bibinfo{pages}{1} (\bibinfo{year}{2017}).

\bibitem[{\citenamefont{Cox et~al.}(2019)\citenamefont{Cox, Maqondo, Kara,
  Milione, Cheng, and Forbes}}]{cox2019hglg}
\bibinfo{author}{\bibfnamefont{M.~A.} \bibnamefont{Cox}},
  \bibinfo{author}{\bibfnamefont{L.}~\bibnamefont{Maqondo}},
  \bibinfo{author}{\bibfnamefont{R.}~\bibnamefont{Kara}},
  \bibinfo{author}{\bibfnamefont{G.}~\bibnamefont{Milione}},
  \bibinfo{author}{\bibfnamefont{L.}~\bibnamefont{Cheng}}, \bibnamefont{and}
  \bibinfo{author}{\bibfnamefont{A.}~\bibnamefont{Forbes}},
  \bibinfo{journal}{Journal of Lightwave Technology}
  \textbf{\bibinfo{volume}{37}}, \bibinfo{pages}{3911} (\bibinfo{year}{2019}),
  ISSN \bibinfo{issn}{0733-8724}, \eprint{1901.07203},
  \urlprefix\url{https://ieeexplore.ieee.org/document/8668467/}.

\bibitem[{\citenamefont{Ndagano
  et~al.}(2017{\natexlab{a}})\citenamefont{Ndagano, Mphuthi, Milione, and
  Forbes}}]{ndagano2017c}
\bibinfo{author}{\bibfnamefont{B.}~\bibnamefont{Ndagano}},
  \bibinfo{author}{\bibfnamefont{N.}~\bibnamefont{Mphuthi}},
  \bibinfo{author}{\bibfnamefont{G.}~\bibnamefont{Milione}}, \bibnamefont{and}
  \bibinfo{author}{\bibfnamefont{A.}~\bibnamefont{Forbes}},
  \bibinfo{journal}{Optics Letters} \textbf{\bibinfo{volume}{42}},
  \bibinfo{pages}{4175} (\bibinfo{year}{2017}{\natexlab{a}}), ISSN
  \bibinfo{issn}{0146-9592},
  \urlprefix\url{https://www.osapublishing.org/abstract.cfm?URI=ol-42-20-4175}.

\bibitem[{\citenamefont{Restuccia et~al.}(2016)\citenamefont{Restuccia,
  Giovannini, Gibson, and Padgett}}]{Restuccia2016}
\bibinfo{author}{\bibfnamefont{S.}~\bibnamefont{Restuccia}},
  \bibinfo{author}{\bibfnamefont{D.}~\bibnamefont{Giovannini}},
  \bibinfo{author}{\bibfnamefont{G.}~\bibnamefont{Gibson}}, \bibnamefont{and}
  \bibinfo{author}{\bibfnamefont{M.}~\bibnamefont{Padgett}},
  \bibinfo{journal}{Optics Express} \textbf{\bibinfo{volume}{24}},
  \bibinfo{pages}{27127} (\bibinfo{year}{2016}), ISSN
  \bibinfo{issn}{1094-4087},
  \urlprefix\url{https://www.osapublishing.org/abstract.cfm?URI=oe-24-24-27127}.

\bibitem[{\citenamefont{Ndagano
  et~al.}(2017{\natexlab{b}})\citenamefont{Ndagano, Mphuthi, Milione, and
  Forbes}}]{ndagano2017comparing}
\bibinfo{author}{\bibfnamefont{B.}~\bibnamefont{Ndagano}},
  \bibinfo{author}{\bibfnamefont{N.}~\bibnamefont{Mphuthi}},
  \bibinfo{author}{\bibfnamefont{G.}~\bibnamefont{Milione}}, \bibnamefont{and}
  \bibinfo{author}{\bibfnamefont{A.}~\bibnamefont{Forbes}},
  \bibinfo{journal}{Optics letters} \textbf{\bibinfo{volume}{42}},
  \bibinfo{pages}{4175} (\bibinfo{year}{2017}{\natexlab{b}}).

\bibitem[{\citenamefont{Trichili et~al.}(2016)\citenamefont{Trichili,
  Rosales-Guzm{\'{a}}n, Dudley, Ndagano, {Ben Salem}, Zghal, and
  Forbes}}]{Trichili2016}
\bibinfo{author}{\bibfnamefont{A.}~\bibnamefont{Trichili}},
  \bibinfo{author}{\bibfnamefont{C.}~\bibnamefont{Rosales-Guzm{\'{a}}n}},
  \bibinfo{author}{\bibfnamefont{A.}~\bibnamefont{Dudley}},
  \bibinfo{author}{\bibfnamefont{B.}~\bibnamefont{Ndagano}},
  \bibinfo{author}{\bibfnamefont{A.}~\bibnamefont{{Ben Salem}}},
  \bibinfo{author}{\bibfnamefont{M.}~\bibnamefont{Zghal}}, \bibnamefont{and}
  \bibinfo{author}{\bibfnamefont{A.}~\bibnamefont{Forbes}},
  \bibinfo{journal}{Scientific Reports} \textbf{\bibinfo{volume}{6}},
  \bibinfo{pages}{27674} (\bibinfo{year}{2016}).

\bibitem[{\citenamefont{Zhao et~al.}(2015)\citenamefont{Zhao, Li, Li, and
  Kahn}}]{zhao2015capacity}
\bibinfo{author}{\bibfnamefont{N.}~\bibnamefont{Zhao}},
  \bibinfo{author}{\bibfnamefont{X.}~\bibnamefont{Li}},
  \bibinfo{author}{\bibfnamefont{G.}~\bibnamefont{Li}}, \bibnamefont{and}
  \bibinfo{author}{\bibfnamefont{J.~M.} \bibnamefont{Kahn}},
  \bibinfo{journal}{Nature photonics} \textbf{\bibinfo{volume}{9}},
  \bibinfo{pages}{822} (\bibinfo{year}{2015}).

\bibitem[{\citenamefont{Zhou et~al.}(2019)\citenamefont{Zhou, Mirhosseini,
  Oliver, Zhao, Rafsanjani, Lavery, Willner, and Boyd}}]{zhou2019using}
\bibinfo{author}{\bibfnamefont{Y.}~\bibnamefont{Zhou}},
  \bibinfo{author}{\bibfnamefont{M.}~\bibnamefont{Mirhosseini}},
  \bibinfo{author}{\bibfnamefont{S.}~\bibnamefont{Oliver}},
  \bibinfo{author}{\bibfnamefont{J.}~\bibnamefont{Zhao}},
  \bibinfo{author}{\bibfnamefont{S.~M.~H.} \bibnamefont{Rafsanjani}},
  \bibinfo{author}{\bibfnamefont{M.~P.} \bibnamefont{Lavery}},
  \bibinfo{author}{\bibfnamefont{A.~E.} \bibnamefont{Willner}},
  \bibnamefont{and} \bibinfo{author}{\bibfnamefont{R.~W.} \bibnamefont{Boyd}},
  \bibinfo{journal}{Optics express} \textbf{\bibinfo{volume}{27}},
  \bibinfo{pages}{10383} (\bibinfo{year}{2019}).

\bibitem[{\citenamefont{Xie et~al.}(2016)\citenamefont{Xie, Ren, Yan, Huang,
  Ahmed, Li, Zhao, Bao, Tur, Ashrafi et~al.}}]{xie2016experimental}
\bibinfo{author}{\bibfnamefont{G.}~\bibnamefont{Xie}},
  \bibinfo{author}{\bibfnamefont{Y.}~\bibnamefont{Ren}},
  \bibinfo{author}{\bibfnamefont{Y.}~\bibnamefont{Yan}},
  \bibinfo{author}{\bibfnamefont{H.}~\bibnamefont{Huang}},
  \bibinfo{author}{\bibfnamefont{N.}~\bibnamefont{Ahmed}},
  \bibinfo{author}{\bibfnamefont{L.}~\bibnamefont{Li}},
  \bibinfo{author}{\bibfnamefont{Z.}~\bibnamefont{Zhao}},
  \bibinfo{author}{\bibfnamefont{C.}~\bibnamefont{Bao}},
  \bibinfo{author}{\bibfnamefont{M.}~\bibnamefont{Tur}},
  \bibinfo{author}{\bibfnamefont{S.}~\bibnamefont{Ashrafi}},
  \bibnamefont{et~al.}, \bibinfo{journal}{Optics letters}
  \textbf{\bibinfo{volume}{41}}, \bibinfo{pages}{3447} (\bibinfo{year}{2016}).

\bibitem[{\citenamefont{Li et~al.}(2017)\citenamefont{Li, Xie, Yan, Ren, Liao,
  Zhao, Ahmed, Wang, Bao, Willner et~al.}}]{li2017power}
\bibinfo{author}{\bibfnamefont{L.}~\bibnamefont{Li}},
  \bibinfo{author}{\bibfnamefont{G.}~\bibnamefont{Xie}},
  \bibinfo{author}{\bibfnamefont{Y.}~\bibnamefont{Yan}},
  \bibinfo{author}{\bibfnamefont{Y.}~\bibnamefont{Ren}},
  \bibinfo{author}{\bibfnamefont{P.}~\bibnamefont{Liao}},
  \bibinfo{author}{\bibfnamefont{Z.}~\bibnamefont{Zhao}},
  \bibinfo{author}{\bibfnamefont{N.}~\bibnamefont{Ahmed}},
  \bibinfo{author}{\bibfnamefont{Z.}~\bibnamefont{Wang}},
  \bibinfo{author}{\bibfnamefont{C.}~\bibnamefont{Bao}},
  \bibinfo{author}{\bibfnamefont{A.~J.} \bibnamefont{Willner}},
  \bibnamefont{et~al.}, \bibinfo{journal}{JOSA B}
  \textbf{\bibinfo{volume}{34}}, \bibinfo{pages}{1} (\bibinfo{year}{2017}).

\bibitem[{\citenamefont{Gu et~al.}(2019)\citenamefont{Gu, Chen, and
  Krenn}}]{krenn2019turbulence}
\bibinfo{author}{\bibfnamefont{X.}~\bibnamefont{Gu}},
  \bibinfo{author}{\bibfnamefont{L.}~\bibnamefont{Chen}}, \bibnamefont{and}
  \bibinfo{author}{\bibfnamefont{M.}~\bibnamefont{Krenn}},
  \bibinfo{journal}{ArXiv} p. \bibinfo{pages}{1906.03581v1}
  (\bibinfo{year}{2019}).

\bibitem[{\citenamefont{Klug et~al.}(2021)\citenamefont{Klug, Nape, and
  Forbes}}]{klug2021orbital}
\bibinfo{author}{\bibfnamefont{A.}~\bibnamefont{Klug}},
  \bibinfo{author}{\bibfnamefont{I.}~\bibnamefont{Nape}}, \bibnamefont{and}
  \bibinfo{author}{\bibfnamefont{A.}~\bibnamefont{Forbes}},
  \bibinfo{journal}{New Journal of Physics} \textbf{\bibinfo{volume}{23}},
  \bibinfo{pages}{093012} (\bibinfo{year}{2021}).

\bibitem[{\citenamefont{Zhan}(2009)}]{zhan2009cylindrical}
\bibinfo{author}{\bibfnamefont{Q.}~\bibnamefont{Zhan}},
  \bibinfo{journal}{Advances in Optics and Photonics}
  \textbf{\bibinfo{volume}{1}}, \bibinfo{pages}{1} (\bibinfo{year}{2009}).

\bibitem[{\citenamefont{Rosales-Guzm{\'a}n
  et~al.}(2018)\citenamefont{Rosales-Guzm{\'a}n, Ndagano, and
  Forbes}}]{rosales2018review}
\bibinfo{author}{\bibfnamefont{C.}~\bibnamefont{Rosales-Guzm{\'a}n}},
  \bibinfo{author}{\bibfnamefont{B.}~\bibnamefont{Ndagano}}, \bibnamefont{and}
  \bibinfo{author}{\bibfnamefont{A.}~\bibnamefont{Forbes}},
  \bibinfo{journal}{Journal of Optics} \textbf{\bibinfo{volume}{20}},
  \bibinfo{pages}{123001} (\bibinfo{year}{2018}).

\bibitem[{\citenamefont{Milione et~al.}(2015)\citenamefont{Milione, Lavery,
  Huang, Ren, Xie, Nguyen, Karimi, Marrucci, Nolan, Alfano
  et~al.}}]{Milione2015d}
\bibinfo{author}{\bibfnamefont{G.}~\bibnamefont{Milione}},
  \bibinfo{author}{\bibfnamefont{M.~P.~J.} \bibnamefont{Lavery}},
  \bibinfo{author}{\bibfnamefont{H.}~\bibnamefont{Huang}},
  \bibinfo{author}{\bibfnamefont{Y.}~\bibnamefont{Ren}},
  \bibinfo{author}{\bibfnamefont{G.}~\bibnamefont{Xie}},
  \bibinfo{author}{\bibfnamefont{T.~A.} \bibnamefont{Nguyen}},
  \bibinfo{author}{\bibfnamefont{E.}~\bibnamefont{Karimi}},
  \bibinfo{author}{\bibfnamefont{L.}~\bibnamefont{Marrucci}},
  \bibinfo{author}{\bibfnamefont{D.~A.} \bibnamefont{Nolan}},
  \bibinfo{author}{\bibfnamefont{R.~R.} \bibnamefont{Alfano}},
  \bibnamefont{et~al.}, \bibinfo{journal}{Optics Letters}
  \textbf{\bibinfo{volume}{40}}, \bibinfo{pages}{1980} (\bibinfo{year}{2015}),
  \urlprefix\url{http://ol.osa.org/abstract.cfm?URI=ol-40-9-1980}.

\bibitem[{\citenamefont{Zhu et~al.}(2021)\citenamefont{Zhu, Janasik, Fyffe,
  Hay, Zhou, Kantor, Winder, Boyd, Leuchs, and Shi}}]{zhu2021compensation}
\bibinfo{author}{\bibfnamefont{Z.}~\bibnamefont{Zhu}},
  \bibinfo{author}{\bibfnamefont{M.}~\bibnamefont{Janasik}},
  \bibinfo{author}{\bibfnamefont{A.}~\bibnamefont{Fyffe}},
  \bibinfo{author}{\bibfnamefont{D.}~\bibnamefont{Hay}},
  \bibinfo{author}{\bibfnamefont{Y.}~\bibnamefont{Zhou}},
  \bibinfo{author}{\bibfnamefont{B.}~\bibnamefont{Kantor}},
  \bibinfo{author}{\bibfnamefont{T.}~\bibnamefont{Winder}},
  \bibinfo{author}{\bibfnamefont{R.~W.} \bibnamefont{Boyd}},
  \bibinfo{author}{\bibfnamefont{G.}~\bibnamefont{Leuchs}}, \bibnamefont{and}
  \bibinfo{author}{\bibfnamefont{Z.}~\bibnamefont{Shi}},
  \bibinfo{journal}{Nature communications} \textbf{\bibinfo{volume}{12}},
  \bibinfo{pages}{1} (\bibinfo{year}{2021}).

\bibitem[{\citenamefont{Ndagano
  et~al.}(2017{\natexlab{c}})\citenamefont{Ndagano, Nape, Perez-Garcia,
  Scholes, Hernandez-Aranda, Konrad, Lavery, and
  Forbes}}]{ndagano2017deterministic}
\bibinfo{author}{\bibfnamefont{B.}~\bibnamefont{Ndagano}},
  \bibinfo{author}{\bibfnamefont{I.}~\bibnamefont{Nape}},
  \bibinfo{author}{\bibfnamefont{B.}~\bibnamefont{Perez-Garcia}},
  \bibinfo{author}{\bibfnamefont{S.}~\bibnamefont{Scholes}},
  \bibinfo{author}{\bibfnamefont{R.~I.} \bibnamefont{Hernandez-Aranda}},
  \bibinfo{author}{\bibfnamefont{T.}~\bibnamefont{Konrad}},
  \bibinfo{author}{\bibfnamefont{M.~P.} \bibnamefont{Lavery}},
  \bibnamefont{and} \bibinfo{author}{\bibfnamefont{A.}~\bibnamefont{Forbes}},
  \bibinfo{journal}{Scientific reports} \textbf{\bibinfo{volume}{7}},
  \bibinfo{pages}{1} (\bibinfo{year}{2017}{\natexlab{c}}).

\bibitem[{\citenamefont{Sit et~al.}(2017)\citenamefont{Sit, Bouchard, Fickler,
  Gagnon-Bischoff, Larocque, Heshami, Elser, Peuntinger, G{\"{u}}nthner, Heim
  et~al.}}]{Sit2017}
\bibinfo{author}{\bibfnamefont{A.}~\bibnamefont{Sit}},
  \bibinfo{author}{\bibfnamefont{F.}~\bibnamefont{Bouchard}},
  \bibinfo{author}{\bibfnamefont{R.}~\bibnamefont{Fickler}},
  \bibinfo{author}{\bibfnamefont{J.}~\bibnamefont{Gagnon-Bischoff}},
  \bibinfo{author}{\bibfnamefont{H.}~\bibnamefont{Larocque}},
  \bibinfo{author}{\bibfnamefont{K.}~\bibnamefont{Heshami}},
  \bibinfo{author}{\bibfnamefont{D.}~\bibnamefont{Elser}},
  \bibinfo{author}{\bibfnamefont{C.}~\bibnamefont{Peuntinger}},
  \bibinfo{author}{\bibfnamefont{K.}~\bibnamefont{G{\"{u}}nthner}},
  \bibinfo{author}{\bibfnamefont{B.}~\bibnamefont{Heim}}, \bibnamefont{et~al.},
  \bibinfo{journal}{Optica} \textbf{\bibinfo{volume}{4}}, \bibinfo{pages}{1006}
  (\bibinfo{year}{2017}).

\bibitem[{\citenamefont{Nape et~al.}(2018)\citenamefont{Nape, Otte, Vall{\'e}s,
  Rosales-Guzm{\'a}n, Cardano, Denz, and Forbes}}]{nape2018self}
\bibinfo{author}{\bibfnamefont{I.}~\bibnamefont{Nape}},
  \bibinfo{author}{\bibfnamefont{E.}~\bibnamefont{Otte}},
  \bibinfo{author}{\bibfnamefont{A.}~\bibnamefont{Vall{\'e}s}},
  \bibinfo{author}{\bibfnamefont{C.}~\bibnamefont{Rosales-Guzm{\'a}n}},
  \bibinfo{author}{\bibfnamefont{F.}~\bibnamefont{Cardano}},
  \bibinfo{author}{\bibfnamefont{C.}~\bibnamefont{Denz}}, \bibnamefont{and}
  \bibinfo{author}{\bibfnamefont{A.}~\bibnamefont{Forbes}},
  \bibinfo{journal}{Optics express} \textbf{\bibinfo{volume}{26}},
  \bibinfo{pages}{26946} (\bibinfo{year}{2018}).

\bibitem[{\citenamefont{Ndagano et~al.}(2018)\citenamefont{Ndagano, Nape, Cox,
  Rosales-Guzman, and Forbes}}]{ndagano2018creation}
\bibinfo{author}{\bibfnamefont{B.}~\bibnamefont{Ndagano}},
  \bibinfo{author}{\bibfnamefont{I.}~\bibnamefont{Nape}},
  \bibinfo{author}{\bibfnamefont{M.~A.} \bibnamefont{Cox}},
  \bibinfo{author}{\bibfnamefont{C.}~\bibnamefont{Rosales-Guzman}},
  \bibnamefont{and} \bibinfo{author}{\bibfnamefont{A.}~\bibnamefont{Forbes}},
  \bibinfo{journal}{Journal of Lightwave Technology}
  \textbf{\bibinfo{volume}{36}}, \bibinfo{pages}{292} (\bibinfo{year}{2018}).

\bibitem[{\citenamefont{Cai et~al.}(2008)\citenamefont{Cai, Lin,
  Eyyubo{\u{g}}lu, and Baykal}}]{cai2008average}
\bibinfo{author}{\bibfnamefont{Y.}~\bibnamefont{Cai}},
  \bibinfo{author}{\bibfnamefont{Q.}~\bibnamefont{Lin}},
  \bibinfo{author}{\bibfnamefont{H.~T.} \bibnamefont{Eyyubo{\u{g}}lu}},
  \bibnamefont{and} \bibinfo{author}{\bibfnamefont{Y.}~\bibnamefont{Baykal}},
  \bibinfo{journal}{Optics express} \textbf{\bibinfo{volume}{16}},
  \bibinfo{pages}{7665} (\bibinfo{year}{2008}).

\bibitem[{\citenamefont{Ji-Xiong et~al.}(2010)\citenamefont{Ji-Xiong, Tao,
  Hui-Chuan, and Cheng-Liang}}]{ji2010propagation}
\bibinfo{author}{\bibfnamefont{P.}~\bibnamefont{Ji-Xiong}},
  \bibinfo{author}{\bibfnamefont{W.}~\bibnamefont{Tao}},
  \bibinfo{author}{\bibfnamefont{L.}~\bibnamefont{Hui-Chuan}},
  \bibnamefont{and}
  \bibinfo{author}{\bibfnamefont{L.}~\bibnamefont{Cheng-Liang}},
  \bibinfo{journal}{Chinese Physics B} \textbf{\bibinfo{volume}{19}},
  \bibinfo{pages}{089201} (\bibinfo{year}{2010}).

\bibitem[{\citenamefont{Wang and Pu}(2008)}]{wang2008propagation}
\bibinfo{author}{\bibfnamefont{T.}~\bibnamefont{Wang}} \bibnamefont{and}
  \bibinfo{author}{\bibfnamefont{J.}~\bibnamefont{Pu}},
  \bibinfo{journal}{Optics communications} \textbf{\bibinfo{volume}{281}},
  \bibinfo{pages}{3617} (\bibinfo{year}{2008}).

\bibitem[{\citenamefont{Cox et~al.}(2016{\natexlab{a}})\citenamefont{Cox,
  Rosales-Guzm\'{a}n, Lavery, Versfeld, and Forbes}}]{Cox:16}
\bibinfo{author}{\bibfnamefont{M.~A.} \bibnamefont{Cox}},
  \bibinfo{author}{\bibfnamefont{C.}~\bibnamefont{Rosales-Guzm\'{a}n}},
  \bibinfo{author}{\bibfnamefont{M.~P.~J.} \bibnamefont{Lavery}},
  \bibinfo{author}{\bibfnamefont{D.~J.} \bibnamefont{Versfeld}},
  \bibnamefont{and} \bibinfo{author}{\bibfnamefont{A.}~\bibnamefont{Forbes}},
  \bibinfo{journal}{Optics Express} \textbf{\bibinfo{volume}{24}},
  \bibinfo{pages}{18105} (\bibinfo{year}{2016}{\natexlab{a}}).

\bibitem[{\citenamefont{Ndagano
  et~al.}(2017{\natexlab{d}})\citenamefont{Ndagano, Mphuthi, Milione, and
  Forbes}}]{Ndagano2017}
\bibinfo{author}{\bibfnamefont{B.}~\bibnamefont{Ndagano}},
  \bibinfo{author}{\bibfnamefont{N.}~\bibnamefont{Mphuthi}},
  \bibinfo{author}{\bibfnamefont{G.}~\bibnamefont{Milione}}, \bibnamefont{and}
  \bibinfo{author}{\bibfnamefont{A.}~\bibnamefont{Forbes}},
  \bibinfo{journal}{Optics Letters} \textbf{\bibinfo{volume}{42}},
  \bibinfo{pages}{4175} (\bibinfo{year}{2017}{\natexlab{d}}), ISSN
  \bibinfo{issn}{0146-9592},
  \urlprefix\url{https://www.osapublishing.org/abstract.cfm?URI=ol-42-20-4175}.

\bibitem[{\citenamefont{Nape et~al.}(2022)\citenamefont{Nape, Singh, Klug,
  Buono, Rosales-Guzman, McWilliam, Franke-Arnold, Kritzinger, Forbes, Dudley
  et~al.}}]{nape2022revealing}
\bibinfo{author}{\bibfnamefont{I.}~\bibnamefont{Nape}},
  \bibinfo{author}{\bibfnamefont{K.}~\bibnamefont{Singh}},
  \bibinfo{author}{\bibfnamefont{A.}~\bibnamefont{Klug}},
  \bibinfo{author}{\bibfnamefont{W.}~\bibnamefont{Buono}},
  \bibinfo{author}{\bibfnamefont{C.}~\bibnamefont{Rosales-Guzman}},
  \bibinfo{author}{\bibfnamefont{A.}~\bibnamefont{McWilliam}},
  \bibinfo{author}{\bibfnamefont{S.}~\bibnamefont{Franke-Arnold}},
  \bibinfo{author}{\bibfnamefont{A.}~\bibnamefont{Kritzinger}},
  \bibinfo{author}{\bibfnamefont{P.}~\bibnamefont{Forbes}},
  \bibinfo{author}{\bibfnamefont{A.}~\bibnamefont{Dudley}},
  \bibnamefont{et~al.}, \bibinfo{journal}{Nature Photonics} pp.
  \bibinfo{pages}{1--9} (\bibinfo{year}{2022}).

\bibitem[{\citenamefont{McLaren et~al.}(2015)\citenamefont{McLaren, Konrad, and
  {Forbes Andrew}}}]{mclaren2015}
\bibinfo{author}{\bibfnamefont{M.}~\bibnamefont{McLaren}},
  \bibinfo{author}{\bibfnamefont{T.}~\bibnamefont{Konrad}}, \bibnamefont{and}
  \bibinfo{author}{\bibnamefont{{Forbes Andrew}}}, \bibinfo{journal}{Physical
  Review A} \textbf{\bibinfo{volume}{92}}, \bibinfo{pages}{23833}
  (\bibinfo{year}{2015}).

\bibitem[{\citenamefont{Ndagano et~al.}(2016)\citenamefont{Ndagano, Sroor,
  McLaren, Rosales-Guzm{\'a}n, and Forbes}}]{ndagano2016beam}
\bibinfo{author}{\bibfnamefont{B.}~\bibnamefont{Ndagano}},
  \bibinfo{author}{\bibfnamefont{H.}~\bibnamefont{Sroor}},
  \bibinfo{author}{\bibfnamefont{M.}~\bibnamefont{McLaren}},
  \bibinfo{author}{\bibfnamefont{C.}~\bibnamefont{Rosales-Guzm{\'a}n}},
  \bibnamefont{and} \bibinfo{author}{\bibfnamefont{A.}~\bibnamefont{Forbes}},
  \bibinfo{journal}{Optics Letters} \textbf{\bibinfo{volume}{41}},
  \bibinfo{pages}{3407} (\bibinfo{year}{2016}).

\bibitem[{\citenamefont{Wootters}(1998)}]{wootters1998entanglement}
\bibinfo{author}{\bibfnamefont{W.~K.} \bibnamefont{Wootters}},
  \bibinfo{journal}{Physical Review Letters} \textbf{\bibinfo{volume}{80}},
  \bibinfo{pages}{2245} (\bibinfo{year}{1998}).

\bibitem[{\citenamefont{Herman and Strugala}(1990)}]{herman1990method}
\bibinfo{author}{\bibfnamefont{B.~J.} \bibnamefont{Herman}} \bibnamefont{and}
  \bibinfo{author}{\bibfnamefont{L.~A.} \bibnamefont{Strugala}}, in
  \emph{\bibinfo{booktitle}{Propagation of High-Energy Laser Beams through
  the\\ Earth's Atmosphere}} (\bibinfo{organization}{International Society for
  Optics and Photonics}, \bibinfo{year}{1990}), vol. \bibinfo{volume}{1221},
  pp. \bibinfo{pages}{183--192}.

\bibitem[{\citenamefont{Rosales-Guzm{\'a}n
  et~al.}(2020)\citenamefont{Rosales-Guzm{\'a}n, Hu, Selyem, Moreno-Acosta,
  Franke-Arnold, Ramos-Garcia, and Forbes}}]{rosales2020polarisation}
\bibinfo{author}{\bibfnamefont{C.}~\bibnamefont{Rosales-Guzm{\'a}n}},
  \bibinfo{author}{\bibfnamefont{X.-B.} \bibnamefont{Hu}},
  \bibinfo{author}{\bibfnamefont{A.}~\bibnamefont{Selyem}},
  \bibinfo{author}{\bibfnamefont{P.}~\bibnamefont{Moreno-Acosta}},
  \bibinfo{author}{\bibfnamefont{S.}~\bibnamefont{Franke-Arnold}},
  \bibinfo{author}{\bibfnamefont{R.}~\bibnamefont{Ramos-Garcia}},
  \bibnamefont{and} \bibinfo{author}{\bibfnamefont{A.}~\bibnamefont{Forbes}},
  \bibinfo{journal}{Scientific Reports} \textbf{\bibinfo{volume}{10}},
  \bibinfo{pages}{1} (\bibinfo{year}{2020}).

\bibitem[{\citenamefont{Scholes et~al.}(2019)\citenamefont{Scholes, Kara,
  Pinnell, Rodr{\'\i}guez-Fajardo, and Forbes}}]{scholes2019structured}
\bibinfo{author}{\bibfnamefont{S.}~\bibnamefont{Scholes}},
  \bibinfo{author}{\bibfnamefont{R.}~\bibnamefont{Kara}},
  \bibinfo{author}{\bibfnamefont{J.}~\bibnamefont{Pinnell}},
  \bibinfo{author}{\bibfnamefont{V.}~\bibnamefont{Rodr{\'\i}guez-Fajardo}},
  \bibnamefont{and} \bibinfo{author}{\bibfnamefont{A.}~\bibnamefont{Forbes}},
  \bibinfo{journal}{Optical Engineering} \textbf{\bibinfo{volume}{59}},
  \bibinfo{pages}{041202} (\bibinfo{year}{2019}).

\bibitem[{\citenamefont{Wang et~al.}(2012{\natexlab{a}})\citenamefont{Wang,
  Yang, Fazal, Ahmed, Yan, Huang, Ren, Yue, Dolinar, Tur
  et~al.}}]{wang2012terabit}
\bibinfo{author}{\bibfnamefont{J.}~\bibnamefont{Wang}},
  \bibinfo{author}{\bibfnamefont{J.-Y.} \bibnamefont{Yang}},
  \bibinfo{author}{\bibfnamefont{I.~M.} \bibnamefont{Fazal}},
  \bibinfo{author}{\bibfnamefont{N.}~\bibnamefont{Ahmed}},
  \bibinfo{author}{\bibfnamefont{Y.}~\bibnamefont{Yan}},
  \bibinfo{author}{\bibfnamefont{H.}~\bibnamefont{Huang}},
  \bibinfo{author}{\bibfnamefont{Y.}~\bibnamefont{Ren}},
  \bibinfo{author}{\bibfnamefont{Y.}~\bibnamefont{Yue}},
  \bibinfo{author}{\bibfnamefont{S.}~\bibnamefont{Dolinar}},
  \bibinfo{author}{\bibfnamefont{M.}~\bibnamefont{Tur}}, \bibnamefont{et~al.},
  \bibinfo{journal}{Nature photonics} \textbf{\bibinfo{volume}{6}},
  \bibinfo{pages}{488} (\bibinfo{year}{2012}{\natexlab{a}}).

\bibitem[{\citenamefont{Cox et~al.}(2016{\natexlab{b}})\citenamefont{Cox,
  Rosales-Guzm{\'a}n, Lavery, Versfeld, and Forbes}}]{cox2016resilience}
\bibinfo{author}{\bibfnamefont{M.~A.} \bibnamefont{Cox}},
  \bibinfo{author}{\bibfnamefont{C.}~\bibnamefont{Rosales-Guzm{\'a}n}},
  \bibinfo{author}{\bibfnamefont{M.~P.} \bibnamefont{Lavery}},
  \bibinfo{author}{\bibfnamefont{D.~J.} \bibnamefont{Versfeld}},
  \bibnamefont{and} \bibinfo{author}{\bibfnamefont{A.}~\bibnamefont{Forbes}},
  \bibinfo{journal}{Optics Express} \textbf{\bibinfo{volume}{24}},
  \bibinfo{pages}{18105} (\bibinfo{year}{2016}{\natexlab{b}}).

\bibitem[{\citenamefont{Zhou et~al.}(2021)\citenamefont{Zhou, Zhao, Braverman,
  Pang, Zhang, Willner, Shi, and Boyd}}]{zhou2021multiprobe}
\bibinfo{author}{\bibfnamefont{Y.}~\bibnamefont{Zhou}},
  \bibinfo{author}{\bibfnamefont{J.}~\bibnamefont{Zhao}},
  \bibinfo{author}{\bibfnamefont{B.}~\bibnamefont{Braverman}},
  \bibinfo{author}{\bibfnamefont{K.}~\bibnamefont{Pang}},
  \bibinfo{author}{\bibfnamefont{R.}~\bibnamefont{Zhang}},
  \bibinfo{author}{\bibfnamefont{A.~E.} \bibnamefont{Willner}},
  \bibinfo{author}{\bibfnamefont{Z.}~\bibnamefont{Shi}}, \bibnamefont{and}
  \bibinfo{author}{\bibfnamefont{R.~W.} \bibnamefont{Boyd}},
  \bibinfo{journal}{Physical Review Applied} \textbf{\bibinfo{volume}{15}},
  \bibinfo{pages}{034011} (\bibinfo{year}{2021}).

\bibitem[{\citenamefont{Wang et~al.}(2012{\natexlab{b}})\citenamefont{Wang,
  Yang, Fazal, Ahmed, Yan, Huang, Ren, Yue, Dolinar, Tur et~al.}}]{Wang2012}
\bibinfo{author}{\bibfnamefont{J.}~\bibnamefont{Wang}},
  \bibinfo{author}{\bibfnamefont{J.-Y.} \bibnamefont{Yang}},
  \bibinfo{author}{\bibfnamefont{I.~M.} \bibnamefont{Fazal}},
  \bibinfo{author}{\bibfnamefont{N.}~\bibnamefont{Ahmed}},
  \bibinfo{author}{\bibfnamefont{Y.}~\bibnamefont{Yan}},
  \bibinfo{author}{\bibfnamefont{H.}~\bibnamefont{Huang}},
  \bibinfo{author}{\bibfnamefont{Y.}~\bibnamefont{Ren}},
  \bibinfo{author}{\bibfnamefont{Y.}~\bibnamefont{Yue}},
  \bibinfo{author}{\bibfnamefont{S.}~\bibnamefont{Dolinar}},
  \bibinfo{author}{\bibfnamefont{M.}~\bibnamefont{Tur}}, \bibnamefont{et~al.},
  \bibinfo{journal}{Nature Photonics} \textbf{\bibinfo{volume}{6}},
  \bibinfo{pages}{488} (\bibinfo{year}{2012}{\natexlab{b}}), ISSN
  \bibinfo{issn}{1749-4885},
  \urlprefix\url{http://dx.doi.org/10.1038/nphoton.2012.138}.

\bibitem[{\citenamefont{Willner and Liu}(2021)}]{willner2021perspective}
\bibinfo{author}{\bibfnamefont{A.~E.} \bibnamefont{Willner}} \bibnamefont{and}
  \bibinfo{author}{\bibfnamefont{C.}~\bibnamefont{Liu}},
  \bibinfo{journal}{Nanophotonics} \textbf{\bibinfo{volume}{10}},
  \bibinfo{pages}{225} (\bibinfo{year}{2021}).

\bibitem[{\citenamefont{Lee}(1979)}]{lee1979binary}
\bibinfo{author}{\bibfnamefont{W.-H.} \bibnamefont{Lee}},
  \bibinfo{journal}{Applied Optics} \textbf{\bibinfo{volume}{18}},
  \bibinfo{pages}{3661} (\bibinfo{year}{1979}).

\bibitem[{\citenamefont{Hu et~al.}(2021)\citenamefont{Hu, Ma, and
  Rosales-Guzm{\'a}n}}]{hu2021high}
\bibinfo{author}{\bibfnamefont{X.-B.} \bibnamefont{Hu}},
  \bibinfo{author}{\bibfnamefont{S.-Y.} \bibnamefont{Ma}}, \bibnamefont{and}
  \bibinfo{author}{\bibfnamefont{C.}~\bibnamefont{Rosales-Guzm{\'a}n}},
  \bibinfo{journal}{Journal of Optics} \textbf{\bibinfo{volume}{23}},
  \bibinfo{pages}{044002} (\bibinfo{year}{2021}).

\bibitem[{\citenamefont{Singh et~al.}(2020)\citenamefont{Singh, Tabebordbar,
  Forbes, and Dudley}}]{singh2020digital}
\bibinfo{author}{\bibfnamefont{K.}~\bibnamefont{Singh}},
  \bibinfo{author}{\bibfnamefont{N.}~\bibnamefont{Tabebordbar}},
  \bibinfo{author}{\bibfnamefont{A.}~\bibnamefont{Forbes}}, \bibnamefont{and}
  \bibinfo{author}{\bibfnamefont{A.}~\bibnamefont{Dudley}},
  \bibinfo{journal}{JOSA A} \textbf{\bibinfo{volume}{37}}, \bibinfo{pages}{C33}
  (\bibinfo{year}{2020}).

\bibitem[{\citenamefont{Selyem et~al.}(2019)\citenamefont{Selyem,
  Rosales-Guzm{\'a}n, Croke, Forbes, and Franke-Arnold}}]{selyem2019basis}
\bibinfo{author}{\bibfnamefont{A.}~\bibnamefont{Selyem}},
  \bibinfo{author}{\bibfnamefont{C.}~\bibnamefont{Rosales-Guzm{\'a}n}},
  \bibinfo{author}{\bibfnamefont{S.}~\bibnamefont{Croke}},
  \bibinfo{author}{\bibfnamefont{A.}~\bibnamefont{Forbes}}, \bibnamefont{and}
  \bibinfo{author}{\bibfnamefont{S.}~\bibnamefont{Franke-Arnold}},
  \bibinfo{journal}{Physical Review A} \textbf{\bibinfo{volume}{100}},
  \bibinfo{pages}{063842} (\bibinfo{year}{2019}).

\bibitem[{\citenamefont{Fried}(1966)}]{Fried1966}
\bibinfo{author}{\bibfnamefont{D.~L.} \bibnamefont{Fried}},
  \bibinfo{journal}{Journal of the Optical Society of America}
  \textbf{\bibinfo{volume}{56}}, \bibinfo{pages}{1372} (\bibinfo{year}{1966}),
  ISSN \bibinfo{issn}{0030-3941},
  \urlprefix\url{http://www.osapublishing.org/viewmedia.cfm?uri=josa-56-10-1372&seq=0&html=true}.

\bibitem[{\citenamefont{Lane et~al.}(1992)\citenamefont{Lane, Glindemann, and
  Dainty}}]{lane1992simulation}
\bibinfo{author}{\bibfnamefont{R.}~\bibnamefont{Lane}},
  \bibinfo{author}{\bibfnamefont{A.}~\bibnamefont{Glindemann}},
  \bibnamefont{and} \bibinfo{author}{\bibfnamefont{J.}~\bibnamefont{Dainty}},
  \bibinfo{journal}{Waves in random media} \textbf{\bibinfo{volume}{2}},
  \bibinfo{pages}{209} (\bibinfo{year}{1992}).

\bibitem[{\citenamefont{Pinnell et~al.}(2020)\citenamefont{Pinnell, Nape,
  Sephton, Cox, Rodr{\'\i}guez-Fajardo, and Forbes}}]{pinnell2020modal}
\bibinfo{author}{\bibfnamefont{J.}~\bibnamefont{Pinnell}},
  \bibinfo{author}{\bibfnamefont{I.}~\bibnamefont{Nape}},
  \bibinfo{author}{\bibfnamefont{B.}~\bibnamefont{Sephton}},
  \bibinfo{author}{\bibfnamefont{M.~A.} \bibnamefont{Cox}},
  \bibinfo{author}{\bibfnamefont{V.}~\bibnamefont{Rodr{\'\i}guez-Fajardo}},
  \bibnamefont{and} \bibinfo{author}{\bibfnamefont{A.}~\bibnamefont{Forbes}},
  \bibinfo{journal}{JOSA A} \textbf{\bibinfo{volume}{37}},
  \bibinfo{pages}{C146} (\bibinfo{year}{2020}).

\bibitem[{\citenamefont{Wootters and Zurek}(1982)}]{wootters1982single}
\bibinfo{author}{\bibfnamefont{W.~K.} \bibnamefont{Wootters}} \bibnamefont{and}
  \bibinfo{author}{\bibfnamefont{W.~H.} \bibnamefont{Zurek}},
  \bibinfo{journal}{Nature} \textbf{\bibinfo{volume}{299}},
  \bibinfo{pages}{802} (\bibinfo{year}{1982}).

\bibitem[{\citenamefont{T{\"o}ppel et~al.}(2013)\citenamefont{T{\"o}ppel,
  Ornigotti, and Aiello}}]{toppel2013goos}
\bibinfo{author}{\bibfnamefont{F.}~\bibnamefont{T{\"o}ppel}},
  \bibinfo{author}{\bibfnamefont{M.}~\bibnamefont{Ornigotti}},
  \bibnamefont{and} \bibinfo{author}{\bibfnamefont{A.}~\bibnamefont{Aiello}},
  \bibinfo{journal}{New Journal of Physics} \textbf{\bibinfo{volume}{15}},
  \bibinfo{pages}{113059} (\bibinfo{year}{2013}).

\end{thebibliography}

\end{document}